%% file: main.tex
\documentclass[sigconf,natbib=true,anonymous=false]{acmart}
\usepackage{enumitem}
\usepackage{tcolorbox}
\usepackage{hyperref}

\newcommand{\tropeheading}[5]{\subsection*{#2 Trope \##1: #3} 
\label{trope:#5}
\textcolor{gray}{-- \textsl{#4} --}}

\newcommand{\xtropeheading}[6]{%
  \item[#6 \hspace{1ex}\textbf{\##1}] \hyperref[trope:#5]{\textbf{#3}: #4 }%
}

\newcommand{\troperef}[2]{%
\hyperref[trope:#1]{#2}%
}

\newcommand{\ld}[1]{} \newcommand{\oz}[1]{} \newcommand{\nc}[1]{}
\newcommand{\pb}[1]{}

\newcommand{\Canva}[0]{Canva}
\newcommand{\Valence}[0]{Valence}

\setcopyright{acmlicensed}
\copyrightyear{2025}
\acmYear{2025} \acmDOI{XXXXXXX.XXXXXXX}
\acmConference[Conference acronym 'XX]{Make sure to enter the correct conference
  title from your rights confirmation email}{Sometime}{Somewhere}
\acmISBN{978-1-4503-XXXX-X/2018/06}

\begin{document}

\title{LLM-Evaluation Tropes: \\
Perspectives on the Validity of LLM-Evaluations}


\author{Laura Dietz}
\affiliation{%
   \institution{University of New Hampshire}
   \country{USA}
}
\author{Oleg Zendel}
\affiliation{%
   \institution{RMIT University}
   \country{Australia}
}
\author{Peter Bailey}
\affiliation{%
   \institution{Canva}
   \country{Australia}
}
\author{Charles Clarke}
\affiliation{%
   \institution{University of Waterloo}
   \country{Canada}
}
\author{Ellese Cotterill}
\affiliation{%
   \institution{Canva}
   \country{Australia}
}
\author{Jeff Dalton}
\affiliation{%
   \institution{University of Edinburgh}
   \country{United Kingdom}
}
\author{Faegheh Hasibi}
\affiliation{%
   \institution{Radboud University}
   \country{Netherlands}
}
\author{Mark Sanderson}
\affiliation{%
   \institution{RMIT University}
   \country{Australia}
}
\author{Nick Craswell}
\affiliation{%
   \institution{Microsoft}
   \country{USA}
}

\renewcommand{\shortauthors}{Dietz, Zendel, Bailey,  Clarke, Cotterill, Dalton, Hasibi, Sanderson, and Craswell}

\begin{abstract}
   \input{abstract}
\end{abstract}


\settopmatter{printfolios=true}

\maketitle









\input{introduction}


\input{llm-tropes}

\input{case-studies}

\input{demonstration}

\input{required-experiments}

\input{conclusion}



\balance
\bibliographystyle{ACM-Reference-Format}
\bibliography{references}

\end{document}

%% file: abstract.tex
Large Language Models (LLMs) are increasingly used to evaluate information
retrieval (IR) systems, generating relevance judgments traditionally made by
human assessors. Recent empirical studies suggest that LLM-based evaluations
often align with human judgments, leading some to suggest that human judges may
no longer be necessary, while others highlight concerns about judgment reliability,
validity, and long-term impact. As IR systems begin incorporating LLM-generated
signals, evaluation outcomes risk becoming self-reinforcing, potentially leading
to misleading conclusions.

This paper examines scenarios where LLM-evaluators may falsely indicate success,
particularly when LLM-based judgments influence both system development and
evaluation. We highlight key risks, including bias reinforcement,
reproducibility challenges, and inconsistencies in assessment methodologies. To
address these concerns, we propose tests to quantify adverse effects, guardrails, and a
collaborative framework for constructing reusable test collections that
integrate LLM judgments responsibly. By providing perspectives from academia and
industry, this work aims to establish best practices for the principled use of
LLMs in IR evaluation.

%% file: introduction.tex
\section{Introduction}

Large Language Models (LLMs) are increasingly used in the evaluation of
information retrieval (IR) systems, generating relevance judgments that were
traditionally the domain of human assessors. Given an information need (or
topic) and a set of documents, assessors determine the relevance of each
document to the topic. A process that forms the foundation of retrieval
evaluation. However, due to the vast number of topic-document pairs, traditional
assessment relies on pooling methods to identify a subset of documents for
judgment. LLMs offer an alternative approach, with the potential to scale
relevance assessments far beyond the limits of human annotation. However, \citet{soboroff2025don} writes: ``Letting the LLM write your truth data handicaps the
evaluation by setting that LLM as a ceiling on performance.'' In this paper we aim for a middle ground by discussing 13 ways in which LLMs can negatively impact the evaluation and how avoid this adverse effect.

\paragraph{Pro.} One of the most impactful advantages of LLM-based evaluation is speed. Unlike
human assessors, who require coordination, training, and extensive annotation
time, LLMs can generate relevance labels almost instantly. This dramatically
lowers the cost of evaluation, making it possible to assess larger datasets,
cover a wider range of retrieval tasks, and conduct evaluations more frequently.
These benefits have led to the rapid adoption of LLMs in large-scale evaluation
pipelines. Microsoft, for example, now uses OpenAI's GPT models for relevance
assessment in Bing~\cite{thomas2024large}. More recently,
\citet{upadhyay2024umbrela} introduced {\sc Umbrela}, an open-source toolkit
based on a similar prompt that uses proprietary LLMs to label unjudged
documents. Its application in a recent TREC task, further reinforces the notion
that human assessors could be replaced~\cite{upadhyay2024llm}. Beyond labeling,
LLMs have been proposed for fully synthetic test collections, where they replace
human users in both query generation and relevance
judgment~\cite{rahmani2024synthetic}.

\paragraph{Con} Although empirical studies demonstrate the effectiveness of LLM-based judgments,
concerns remain regarding their reliability, validity, and long-term
implications for IR evaluation \cite{faggioli2023perspectives}. The increasing reliance on LLMs for test
collection creation raises fundamental questions about reproducibility. At the
same time traditional pooled judgment methodologies are becoming impractical for
assessing generative and multi-modal systems. Furthermore, there is no consensus
on how to mitigate the risks associated with these models, including biases,
inconsistencies, and potential vulnerabilities such as susceptibility to query
stuffing~\cite{Alaofi2024}. Without a shared principled framework for
responsible adoption, the unchecked use of LLMs in evaluation studies may lead
to misleading conclusions, where prior work is cited without appropriate
consideration of its limitations.

Below we outline our contributions by specifying  questions that are addressed in this paper.


\subsection{Can I Use LLM-judgments in My Next Research Paper?}



\paragraph{Contributions} This paper explores validity challenges posed by using LLMs as a evaluator, aims to
codify the best practices for ensuring that LLM-based evaluation remains a valid
experimentation approach for IR research.  Maintaining scientific rigor requires
identifying and recognizing the risks,  implementing guardrails, and continuously
confirming their validity with human judgments.

Based on many discussions and reviewing the literature, we establish a taxonomy
of reoccurring issues with LLM evaluations, which we refer to as LLM Evaluation
Tropes (Figure \ref{fig:tropes}).  To this end, we outline guardrails that
should be applied when using LLM-based evaluation for research
publications.\footnote{We hope our recommendations offer a balanced compromise
that satisfies both authors and reviewers.} We underscore the truthfulness and
relevance of these tropes by case studies from academia and industry. This paper
is co-authored by a team of established international researchers from who hold a spectrum of viewpoints.




\subsection{What is a Valid Experiment?}

We take \emph{valid} LLM evaluators to mean evaluators whose
measurements align with human intuition and quantify the utility for real people. We follow the guidance from
\citet{spark1975report}:

\begin{quote}
    ``It is apparent in particular that it is most important that the ideal
    collection(s) should be a means of relating \textbf{valid abstract studies}
    of information retrieval and those of operational systems and user
    behaviour.''
\end{quote}

In industry, the effectiveness of systems is typically validated through A/B
testing and manual assessments. In academia, evaluation
helps to determine which approaches constitute research advances and should be
submitted to conferences and shared tasks. In both industry and academia, it is
essential to obtain quantitative quality measures that credibly reflect
relevance while mitigating unintentional biases, such as test data leakage,
which may compromise the validity of drawn conclusions.

In this paper, we discuss conditions under which the use of LLM evaluators may (inadvertently) threaten
the validity of the experiment.

\input{trope-overview}

\subsection{Why Does the Old Evaluation Paradigm No Longer Apply?}

The long standing and widely accepted testing paradigm, as practiced in
TREC,\footnote{\url{https://trec.nist.gov/}}
CLEF,\footnote{\url{https://www.clef-initiative.eu/}}
FIRE,\footnote{\url{https://fire.irsi.org.in}} and
NTCIR,\footnote{\url{https://research.nii.ac.jp/ntcir/index-en.html}} assumes
that all effective IR systems will identify a similar set of relevant documents.
To approximate a comprehensive relevance set, pooling techniques select
top-ranked documents from multiple systems for manual
assessment~\cite{Voorhees2019}.

While this approach remains foundational, it has diverged from the original
Cranfield experiment~\cite{cleverdon1967}, in which \emph{all} documents were
judged rather than a limited subset of potentially relevant ones. Since
Cranfield's inception in 1967, the scale of document collections, including the
Web, has expanded exponentially, making exhaustive manual assessment infeasible.
Moreover, when synthesis or generation is introduced into IR systems, many of
the existing assumptions collapse. Instead of a static search engine result page
(SERP) containing a fixed set of documents, generative models produce
paraphrased or alternative responses, many of which may be equally valid.
Assessing all such variations manually is impractical. At the same time, a
single word change can render a relevant response to be non-relevant.

Even without generative models, the traditional evaluation setup faces
limitations. Retrieval systems that are not included in the original pooling process
often return unjudged documents, leading to incomplete evaluations. This issue
is particularly acute for newer retrieval paradigms -- such as dense retrieval,
neural re-ranking, and query reformulation -- which retrieve documents in ways
that diverge from those originally used to construct the test
collections~\cite{Roberts2020}. As a result, many established test collections
are becoming less useful for evaluating newer retrieval models, especially when
these models differ substantially from those used in the pooling process. 

\bigskip
To
address these challenges, the IR community has long relied on approximations to
make the evaluation tractable, while continuously refining evaluation
methodologies to uphold
validity~\cite{Moffat2008,Frobe2023,Lu2017,Roitero2022,Moffat2015,Sakai2021}.
The extensive body of methodological research developed over the years provides
valuable insights for designing guardrails and validation techniques in emerging
evaluation paradigms.

We call for the development of novel evaluation paradigms that uphold the validity of experimental results, and for reusable test collection construction methods that minimize the need for additional human assessments.

\subsection{Outline}
We start by exploring different ways in which the use of LLMs in evaluation can inadvertently negatively impact the validity of the evaluation.  
Section  \ref{sec:usecases} supports this with case studies from two companies who identified and overcame validity issues in their experimentation. We include a demonstration of how circularity can arise with data from a recent TREC task in Section \ref{sec:demonstration} and suggest a annual effort to cooperatively build test collections with recent systems, evaluators, and content modifiers in Section \ref{sec:coopetition},  before concluding the paper.


%% file: trope-overview.tex
\newcommand{\circsym}{$\circlearrowright$}
\newcommand{\variety}{$\cong$}
\newcommand{\goodhart}{$\heartsuit$} 

\begin{figure}
\caption{LLM-evaluation tropes that can lead to invalid conclusions about evaluation, systems, and the efficacy of human judges that oversee the process. Overarching patterns are circularity \circsym, Goodhart's law \goodhart, and loss of variety \variety.
\label{fig:tropes}}

\begin{tcolorbox}[colframe=black, colback=white, coltitle=black, width=\columnwidth, boxrule=0.2mm, arc=3mm,  halign=left]
\flushleft

\textbf{Evaluation Tropes:}
\begin{itemize}
\xtropeheading{1}{Eval}{Circularity}{Leaking the evaluation signal into the IR system.}{circularity}{\circsym}
\xtropeheading{2}{Eval}{LLM-Eval as a Ranker}{Using the same approach in the system and the evaluation.}{eval_as_ranker}{\circsym}
\xtropeheading{3}{Eval}{LLM Narcissism}{LLMs prefer text from their own model.}{narcissism}{\goodhart}
\xtropeheading{4}{Eval}{Loss of Variety of Opinion}{When all judges think alike.}{diversity}{\variety}
\end{itemize}

\textbf{Meta-Evaluation Tropes:}
\begin{itemize}

\xtropeheading{5}{Meta-Eval}{Ignored Label Correlation}{
When human and LLM judges disagree on relevance labels.}{correlation}{\goodhart}
\xtropeheading{6}{Meta-Eval}{Old Systems}{Evaluators need to identify the best systems of the future.}{old-systems}{\variety}
\xtropeheading{7}{Meta-Eval}{LLM Evolution}{LLMs are not static; they can improve or degrade over time.}{evolution}{}
\end{itemize}

\textbf{System Tropes:}
\begin{itemize}

\xtropeheading{8}{System}{Test Set Leak}{LLMs trained on test collections create the illusion of quality.}{test-set-leak}{\goodhart}
\xtropeheading{9}{System}{Self-Training Collapse}{Concept drift from training LLMs on LLM output.}{concept-drift}{\circsym}
\xtropeheading{10}{System}{Goodhart's Overfitting}{IR systems that are ``trained for the test''.}{synthetic}{\goodhart \!\!}
\xtropeheading{11}{System}{Adversarial Threats}{Bad actors want to manipulate the systems and evaluation.}{adversarial}{\goodhart \!\!}
\end{itemize}

\textbf{Judge Tropes:}
\begin{itemize}

\xtropeheading{12}{Judge}{Rubber-Stamp Effect}{Lack of critical
    oversight when humans blindly trust LLM labels.}{rubber}{\variety \!\!}
\xtropeheading{13}{Judge}{Black-box Labeling}{When relevance is complex, labels may be difficult to interpret.}{blackbox}{\hspace{1ex}}
\xtropeheading{14}{Judge}{Predictable Secrets}{When human data can be guessed by an LLM.}{guess}{\circsym \!\!}

\end{itemize}
\end{tcolorbox}
\vspace{-2em}
\end{figure}

%% file: llm-tropes.tex
\section{LLM-Eval Tropes and Guardrails}
\label{sec:tropes}

During system development, various forms of Goodhart's Law \cite{goodhart1975problems} and test data leakage may compromise evaluation integrity. A critical hazard is \emph{circularity}, a feedback loop in which evaluator assumptions and system design decisions reinforce one another, distorting evaluation outcomes. As a result, evaluation metrics may no longer reflect human preferences under realistic conditions, thereby invalidating the evaluation paradigm.

We present a taxonomy of recurring tropes observed in LLM-based evaluations (see Figure~\ref{fig:tropes} for an overview). These tropes can negatively affect different stages of the evaluation process, and we discuss several types that pose particular challenges to evaluation validity:

\begin{enumerate} 
    \item \textbf{Eval}: Tropes that lead to misleading or incorrect evaluation measurements. 
    \item \textbf{Meta-Eval}: Tropes that give the false impression of high evaluation quality or reliability. 
    \item \textbf{System}: Tropes that cause an IR system to perform poorly or unreliably in real-world scenarios.
    \item \textbf{Judge}: Tropes that inadvertently undermine the effectiveness of human judgment in the evaluation process. 
\end{enumerate}

Below, we examine these common trope patterns, highlighting their pitfalls, and
propose guardrails to mitigate their shortcomings.

\subsection{Evaluation Tropes}

We begin by describing a set of common evaluation tropes that can undermine the validity of LLM-based evaluation systems. These issues arise when the design or application of the evaluation process produces misleading metrics, circular validation, or inflated estimates of system performance.

\tropeheading{1}{Eval}{Circularity}{Leaking the evaluation signal into the IR system.}{circularity}

When the output of an LLM evaluator influences the design or training of the IR system, the evaluation risks becoming self-reinforcing rather than genuinely informative \citep{alaofi2024generative, clarke2024llm, wei2024rethinking, balog2025rankers}.

Even when system developers and LLM evaluators work independently,
unintentional contamination can occur \cite{faggioli2023perspectives}. A system may unknowingly integrate
aspects of an LLM evaluator's methodology, either through training data,
algorithmic choices, or shared heuristics. This accidental feedback loop can
result in inflated performance in LLM-based evaluations without corresponding gains for real-world applications
\citep{singh2024evaluation,sainz2023nlp,roberts2023data,balog2025rankers}.

A related concern is that repeated circularity, i.e., when systems are  trained and evaluated with
signals from the same LLM evaluator, may result in effects of the \troperef{concept-drift}{Model
Collapse trope}~\cite{shumailov2024ai}. 

\paragraph{Quantify effect}
The impact of this trope can be assessed by comparing against a manual evaluation paradigm and measuring divergence in leaderboard rankings -- particularly for systems that may have been influenced by evaluation signal leakage. We present one such study in Section \ref{sec:demonstration}.

\paragraph{Guardrail}
Involving human annotators offers independent  against self-reinforcing evaluation loops. Care must be taken to steer clear of the \troperef{guess}{Predictable Secret} trope.

\tropeheading{2}{Eval}{LLM-Eval as a Ranker}{Using the same approach in the system and the evaluation.}{eval_as_ranker}

This trope refers to a special case of circularity that arises when the same LLM
evaluator is used both within the ranking system and as the evaluation metric. This trope effects from any form of self-refinement procedure \cite{xu2024pride}. In the context of IR, this arises when the same procedure generates both the rank scores and the evaluation scores, the result is a superficial alignment that yields an artificially perfect evaluation—despite the system potentially performing poorly under human judgment \citep{gao2024llmenhanced,clarke2024llm,balog2025rankers}.

A similar failure mode can be illustrated with BM25: if the top ten documents retrieved by BM25 were assumed to define ground-truth relevance, then a BM25 ranker would trivially achieve perfect P@10. Clearly, no one would accept such a circular and and invalid evaluation paradigm -- except in limited contexts such as improving the efficiency of a system \cite{clarke2016assessing}.

We recognize that system developers will want to include notions of
LLM-evaluation in their system. In Section~\ref{sec:demonstration}  we examine this scenario in the context of the TREC RAG 2024 track, which employed the {\sc Umbrela} LLM evaluator. 
We show that reranking submitted runs using {\sc Umbrela} improves performance under manual assessment -- a valid and actionable finding that supports system-side use of LLM evaluators.

However, evaluating such reranked systems using the same {\sc Umbrela} metric introduces circularity and \textbf{leads to invalid evaluation outcomes}.  In our analysis, this reuse of the evaluator causes significant divergence from human assessments: human and LLM evaluators disagree on the relative quality of 18\% of system pairs--more than twice as many as found on original systems. Moreover, while twelve systems score above 0.95 in {\sc Umbrela}-NDCG, their manual NDCG scores range from only 0.68 to 0.72.

\paragraph{Quantify effect}
This effect can be quantified by repeating the analysis in Section~\ref{sec:demonstration}, directly comparing system rankings under LLM and human evaluation.

\paragraph{Guardrail}
Avoid using an LLM evaluation procedure if the same (or closely related) procedure may be embedded within the system under evaluation. 

\tropeheading{3}{Eval}{LLM Narcissism}{LLMs prefer text from their own model.}{narcissism}

Being language models, LLM evaluators tend to assign higher scores to text that aligns closely with their own generation patterns—effectively equating textual quality with per-token likelihood. 
This leads to a preference for outputs produced by the same model family. For instance, GPT-4 may systematically favor responses generated by GPT-4-based systems, even when human assessors detect no meaningful quality difference \cite{panickssery2024llm,liu-etal-2023-g,liu2024llms,xu2024pride}. 
This results in distorted system rankings and compromises the validity of the evaluation outcomes.

Models may be optimized to align with the LLM's biases rather than real-world
relevance assessments, undermining the credibility of the evaluation.

\paragraph{Quantify effect}
The experimental protocol from \citet{liu-etal-2023-g} can quantify this effect. It involves recording the LLM versions used in both systems and evaluators, and analyzing how often evaluators favor systems built with the same underlying model.

\paragraph{Guardrail}
One mitigation strategy is to reserve a specific LLM (or family of LLMs) exclusively for evaluation purposes, ensuring it is not used in any system under test. However, due to overlapping training corpora across models, this bias may still persist. A more robust alternative is to involve multiple LLMs in the evaluation and aggregate relevance judgments using majority voting -- while omitting the vote of any evaluator that shares lineage with the system under consideration.

\begin{figure}
    \centering
    \includegraphics[width=1\linewidth]{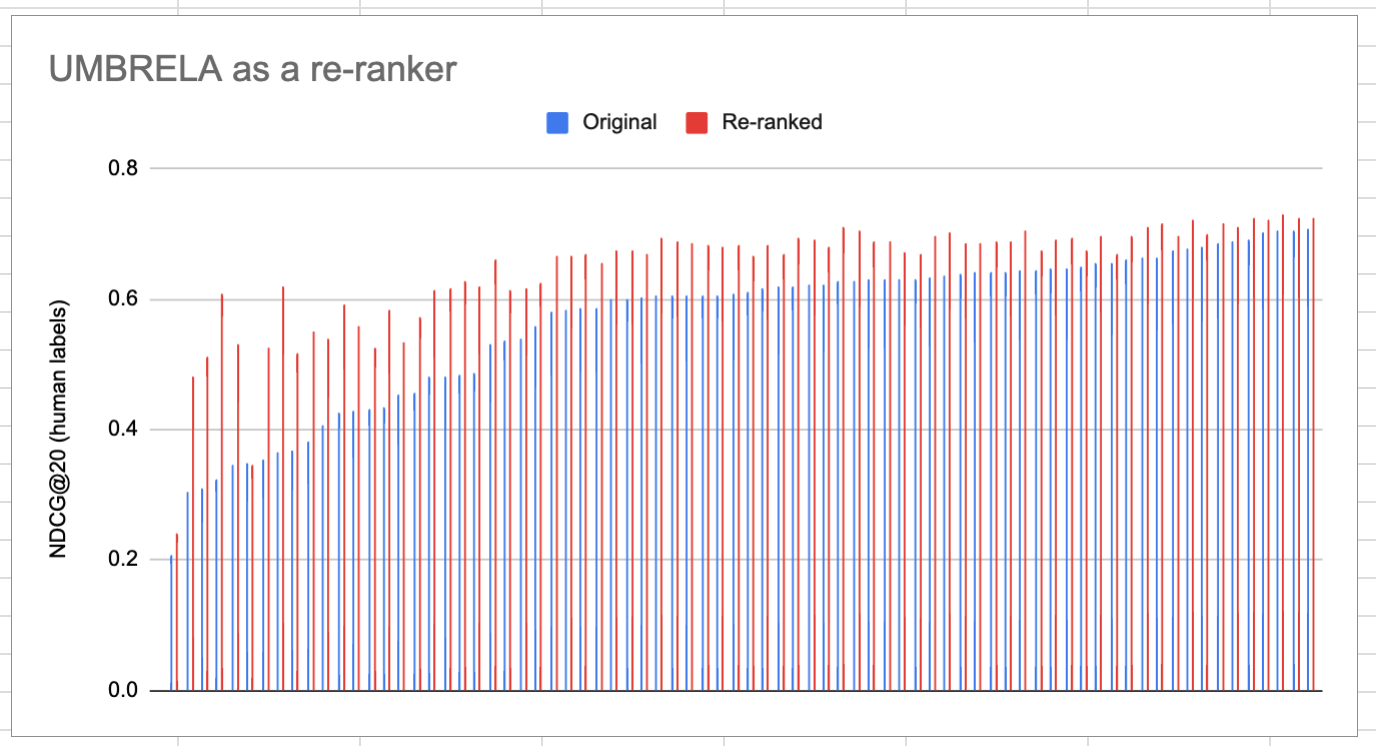}
    \caption{Reranking with an LLM evaluator ({\sc Umbrela}) improves performance under human relevance labels. This plot compares the original and reranked versions of all TREC RAG 24 systems based on manual assessment.}

    \label{fig:rag24-llmranker}
\end{figure}

\tropeheading{4}{Eval}{Loss of Variety of Opinion}{When all judges think alike.}{diversity}

LLM-based evaluations risk homogenizing judgment. Prior work has shown that LLMs can exhibit gender and cultural biases, often reinforcing dominant perspectives while penalizing creative, diverse, or unconventional—yet valid—outputs \citep{padmakumar2023writing,bhatt2024extrinsic}. In contrast, human assessors are better equipped to recognize nuance, novelty, and contextual diversity, which LLMs frequently overlook \citep{si2024evaluating}.

More fundamentally, when LLMs define what is relevant across the board, they implicitly set a ceiling for what systems can achieve. This can penalize systems that offer innovative or non-standard responses that fall outside the LLM's implicit norms \citep{soboroff2025don}.

\paragraph{Quantify effect}
This trope’s impact can only be assessed through independent evaluations involving human judges from diverse socio-cultural backgrounds.

\paragraph{Guardrail}
While human annotation workflows can be designed to ensure a variety of perspectives, achieving this with LLMs is far more difficult. Persona-based prompting strategies \citep{tseng2024two} have been proposed as a mitigation, but emerging evidence highlights their limitations \citep{li2025llm}. We recommend rigorous quantification of this effect before relying on such methods in high-stakes evaluation.

\subsection{Meta-Evaluation Tropes}

The quality of different LLM-judge approaches is often validated through meta-evaluation -- a paradigm that measures how well LLM judgments reproduce either manually created relevance labels or leaderboard rankings under an official evaluation metric \cite{zhou2025evaluating,faggioli2023perspectives}. However, such meta-evaluations can foster a false sense of reliability or progress, masking deeper issues in metrics, methodology, or evaluator behavior.

\tropeheading{5}{Meta-Eval}{Ignored Label Correlation}{
When human and LLM judges disagree on relevance labels.}{correlation}

Meta-evaluations of LLM-based judges often rely on measuring the correlation between system rankings (or document rankings for individual  queries) based on an evaluation metric (e.g., using NDCG or DCG) using  manual and LLM-based labels  \citep{upadhyay2024umbrela,Oosterhuis:2024:RCI}. However, this high-level agreement can mask important differences at the level of individual judgments.

For example, in the context of conversational systems, \citet{mehri-2020-usr} demonstrate that even when system-level Spearman correlation is perfect (i.e., $\rho = 1$), agreement on individual relevance labels can vary widely -- from as low as 0.12 to 0.61 -- depending on the underlying metric. This highlights the risk of relying solely on system-level comparisons.

To establish that LLM-generated judgments are robust and future-proof -- and do not artificially constrain the performance of emerging systems -- agreement should be assessed directly at the label level, i.e., for each query-document pair.

\paragraph{Quantify effect}
Measure correlation between human and LLM-generated relevance labels directly, to assess the reliability of LLM judgments at the label level.

\paragraph{Guardrail}
Incorporating label-level agreement analysis alongside system-level metrics ensures that inconsistencies or biases in LLM evaluations are not overlooked, providing a more complete view of evaluator reliability.

\tropeheading{6}{Meta-Eval}{Old Systems}{Evaluators need to identify the best systems of the future.}{old-systems}

The primary goal of evaluation is to identify the next generation of state-of-the-art systems. Accordingly, meta-evaluations of LLM-based judges aim to show that these evaluators can correctly identify the best-performing systems. However, this claim is often tested on \emph{legacy systems} -- those that were state-of-the-art at the time the test collection was created.

As new IR paradigms emerge, they are rarely reflected in existing test collections.
As a result, a meta-evaluation on a dataset can only confirm whether the LLM evaluator recognizes high-performing systems that era.

Yet such studies are frequently used to argue that LLM evaluators will \emph{also} be effective for future systems. This assumption remains untested for future systems, which are expected to differ significantly. Such systems are likely to employ LLMs more extensively, integrate higher-quality models, or adopt innovations that differ significantly from past approaches. There is a real danger that LLM evaluators -- especially those themselves based on outdated LLMs -- may fail to recognize these future breakthroughs.

\paragraph{Quantify effect}

A simple change in community practice is to collect implementations of the
recent IR systems and release an expanded judgment pools  (e.g. as suggested in
Section \ref{sec:coopetition}). 
By repeating meta-evaluations on these new systems using test collection artifacts, one can assess whether the LLM evaluator still identifies best performing systems.

\paragraph{Guardrail}
Older TREC collections remain relevant, because of available manual runs \cite{voorhees2022can}. In addition, the community should regularly collect recent system implementations, expand the relevance pool, and re-run meta-evaluations to detect and mitigate the effects of this trope.

\tropeheading{7}{Meta-Eval}{LLM Evolution}{LLMs are not static; they can improve or degrade over time.}{evolution}

Meta-evaluations of LLM-based judges often rely on a single prompt or a single LLM family, despite the wide variety of models available. Crucially, LLMs are not static -- model behavior evolves over time as new versions are released \citep{chen2023chatgpt}. Future iterations of an LLM may judge relevance differently than earlier ones, introducing inconsistencies in longitudinal evaluations. This drift becomes especially problematic when newer models are trained on data that includes outputs from earlier versions, potentially leading to feedback loops and self-training collapse \cite{shumailov2024ai}.

These issues are compounded by the fact that LLM providers may seamlessly retire older versions or update models without notice.\footnote{\url{https://openai.com/index/gpt-4-api-general-availability/}} This makes it difficult -- or even impossible -- to reproduce prior evaluation findings using the same version of the evaluator.

\paragraph{Quantify effect}
To track the impact of model evolution, meta-evaluations should be periodically repeated using updated LLM versions. Key indicators of behavioral drift include changes in the ranking of top systems, inconsistencies in relevance labels compared to human judgments, and increased variability in the labeling of previously unjudged documents \cite{arabzadeh2025human,abbasiantaeb2024can}.

\paragraph{Guardrail}
Because access to specific model versions cannot be guaranteed over time, LLM-based evaluation methods must be continually re-validated. In Section \ref{sec:coopetition}, we recommend that the community adopt a recurring meta-evaluation protocol to ensure the ongoing reliability and relevance of LLM-based evaluators.  

\subsection{System Tropes}

Next, we examine tropes that degrade IR system performance as a result of reliance on artifacts such as LLM-generated relevance labels. While synthetic data and automated evaluators can improve scalability, their improper use can introduce systemic biases and encourage overfitting to unreliable or unstable evaluation signals.

\tropeheading{8}{System}{Test Set Leak}{LLMs trained on test collections create the illusion of quality.}{test-set-leak}

Some LLMs are trained on publicly available test collections used in IR evaluation. This contaminates evaluation outcomes by inflating the performance of systems that incorporate such LLMs, creating the illusion of high accuracy that fails to generalize to real-world scenarios \citep{zhou2023dont}.

There are, however, legitimate reasons to train an LLM on relevance labels, for example, when developing an LLM evaluator specifically designed to support assessment. Prior work, such as the Autotar evaluation framework \citep{cormack2015autonomy,cormack2016scalability}, demonstrates that targeted training can yield valid and scalable evaluation systems.

Nevertheless, if such test collections are also used in meta-evaluation, training-induced memorization can create a misleading appearance of alignment between LLM and human judgments—undermining the credibility of the evaluation approach \cite{deng2024investigating,xu2024benchmark}.

\paragraph{Quantify effect}
After collecting fresh manual relevance labels on new topics for the task, a drop in performance on fresh topics would signal potential overfitting or memorization. The protocol of \citet{bordt2024elephants}, developed in the context of table learning, provides a useful template for quantifying this effect.

\paragraph{Guardrail}
Avoid conducting IR research on test collections likely to have been included in LLM training data, as this risks measuring memorization rather than generalization. Regularly collect fresh human relevance judgments on new topics to track performance drift. Because most LLM training corpora are not disclosed, a trusted entity should maintain a hidden subset of evaluation topics to safeguard against future test set leakage.

\tropeheading{9}{System}{Self-Training Collapse}{Concept drift from training LLMs on LLM output.}{concept-drift}

The increasing use of LLM-generated content as training data for other LLMs raises serious concerns about concept drift and long-term quality degradation \citep{yeo2024selftraining}. Rather than fostering diversity or nuance, repeated training on synthetic outputs may entrench biases and amplify systematic errors \citep{gerstgrasser2024model,dohmatob2024demystified, shumailov2023curse,li2025preference}.

In the context of IR, this phenomenon arises when LLM-based evaluators are used to generate synthetic training data for IR systems. The problem compounds when the outputs of these synthetically trained systems are then used to fine-tune the next generation of LLM evaluators, forming a recursive feedback loop. This recursive co-training process can amplify subtle biases and lead to concept drift -- and, ultimately, model collapse~\cite{shumailov2024ai}.  This is a concrete manifestation of unintended circularity during system development, wherein models achieve high training or evaluation scores but fail to generalize in real-world scenarios.


\paragraph{Quantify effect}
This effect can be quantified  \citep{shumailov2023curse} by tracking evaluation performance on a fixed set of held-out, human-labeled data across multiple rounds of recursive training. A consistent decline in agreement with manual judgments would signal the onset of model collapse.

\paragraph{Guardrail}
One should adopt guardrails from  Reinforcement Learning from AI Feedback
\citep{dohmatob2024demystified} and  generative AI  \citep{shumailov2023curse}
to detect when systems are degrading.

To avoid inadvertently exercising this trope, training data should be released
with proper documentation of how training data was obtained, and to which extent
LLMs were used in the generation.

\tropeheading{10}{System}{Goodhart's Overfitting }{IR systems that are ``trained for the test''.}{synthetic}

When a metric becomes a target, it ceases to be a reliable measure of success \citep{goodhart1975problems}. In IR, this effect arises when systems overfit to artifacts of a specific LLM-based evaluator, rather than optimizing for genuine user satisfaction. Industry experience \citep{thomas2024matters,zhang2020models} has shown that systems tuned solely for a single metric (such as high NDCG) may underperform on user-centered outcomes, including click-through rate, dwell time, and robustness to spam.

We anticipate similar issues for systems optimized against a single LLM evaluator, e.g., {\sc Umbrela}, which may achieve high scores under that metric while degrading on alternative evaluation measures or in real-world effectiveness.

\paragraph{Quantify effect}
\citet{ailem2024examining} proposes a protocol for quantifying overfitting by using a range of evaluation metrics.

\paragraph{Guardrail}
To mitigate overfitting in this trope, evaluations should include multiple complementary metrics, along with manual relevance labels. This approach helps detect when systems are narrowly optimizing for a specific LLM-derived signal while failing to generalize across other important evaluation dimensions \citep{zhang2023constructing,zhang2020better,scheltema2024recommender}.

\tropeheading{11}{System}{Adversarial Threats}{Bad actors want to manipulate the systems and evaluation.}{adversarial}

Adversarial behavior is an increasing concern \citep{kurland2022,basat2015} as LLMs become central to both retrieval and evaluation pipelines. These models are susceptible to manipulation, particularly via the \troperef{narcissism}{LLM Narcisissm Trope}, which can be exploited for search engine optimization (SEO).

Recent studies demonstrate that LLMs can be guided to rewrite content to improve evaluation scores \citep{bardas2025,wang2024}. Such techniques can distort rankings, spread misinformation, or amplify propaganda. System developers, aware of evaluation setups, may optimize their outputs to align with known evaluator biases -- effectively training for the test. This risk increases when evaluation models and prompts are publicly disclosed, enabling targeted reverse-engineering.


LLMs can also be deceived into labeling irrelevant documents as relevant using simple adversarial attacks \citep{Alaofi2024,shi2024optimization,parry2024analyzing}. These vulnerabilities threaten the integrity of evaluation pipelines and call into question the trustworthiness and reliability of LLM-based assessments.

\paragraph{Quantify effect}
This effect can be measured by analyzing performance changes when outputs are explicitly optimized for a specific LLM evaluator, prompt, or configuration. Comparative studies help estimate to which extent evaluation scores are inflated by evaluation-aware tuning.

\paragraph{Guardrail}

We advocate developing adversarial test inputs, e.g., targeted content rewrites, to assess the resilience of evaluation metrics under manipulation (cf. Section \ref{sec:coopetition}).

To reduce vulnerability, evaluation campaigns (e.g., TREC) should avoid exposing evaluator identities and prompt designs. Blind evaluation setups, where system developers are unaware of the specific LLM and prompt, can reduce gaming. Rotating or ensembling multiple evaluators adds further robustness. Where feasible, human judgments should remain part of the evaluation loop to validate and audit automated assessments.

\subsection{Judge Tropes}

A common solution to many evaluation and system tropes is to incorporate human judges into the evaluation process. Rather than relying solely on pristine manual judgments, many current approaches involve a collaboration between human assessors and LLMs to generate relevance labels. However, this hybrid setup introduces new risks: subtle forms of bias that can arise during human verification. We refer to these as ``judge tropes''.

While human involvement is often viewed as the gold standard, misalignment between task design, instructions, or expectations can inadvertently render human judgments ineffective -- or even misleading. These issues can compromise the integrity of relevance assessments and, in severe cases, invalidate experimental findings. Biases may stem from overreliance on LLM outputs, cognitive fatigue, or inadequate oversight -- emphasizing the need for robust guardrails and diverse, well-calibrated evaluation protocols.

\tropeheading{12}{Judge}{Rubber-Stamp Effect}{Lack of critical
    oversight when humans blindly trust LLM labels.}{rubber}

Experimental studies show that when human assessors are shown LLM-generated answers before making their own judgments, they are significantly more likely to conform to the model’s assessment -- even when it is demonstrably incorrect \citep{fok2024search,distrust2023,krugel2021zombies,steyvers2025large}. In psychology this is known as Ash conformity experiments \cite{asch1958effects,griffin2023susceptibility}. Moreover, as assessor fatigue and task repetition set in, human verification of LLM-generated labels often turns into passive agreement, driven more by trust than by critical scrutiny.

This creates a feedback loop: despite involving human judges, evaluations increasingly mirror LLM outputs -- even when those outputs diverge from human intuition or real-world utility.

\paragraph{Quantify effect}
This effect can be measured by comparing outcomes under fully manual relevance labels versus human-verified LLM labels. Divergence in label quality or ranking decisions will help quantify the degree of automation bias introduced.

\paragraph{Guardrail} 
To counteract this effect, we draw inspiration from vigilance protocols in security contexts \citep{claypoole2019effects}. We propose embedding vigilance tests in the annotation workflow—rewarding annotators for identifying errors in LLM outputs. Randomly flipped or adversarial labels can be inserted to test whether annotators are critically engaged. If assessors fail to flag introduced errors, this signals a breakdown in oversight and provides a measurable indicator of rubber-stamping behavior.

\tropeheading{13}{Judge}{Black-box Labeling}{When relevance is complex, labels may be difficult to interpret.}{blackbox}

Relevance labels are often used to represent complex judgments in a simplified form. Whether assigned by humans or LLMs, it can be difficult to determine why a particular passage received a given label -- especially when the decision is based on multiple, opaque criteria. This challenge is exacerbated when LLMs provide relevance judgments without clear or trustworthy rationales, increasing the risk of uncritical acceptance by human verifiers \citep{liao2023ai,ji2024detecting}.

Lack of transparency in LLM-generated relevance labels is a concern.
Although LLM evaluators can generate explanations alongside labels, these rationales may themselves be flawed and must be critically assessed ~\cite{shumailov2024ai} while avoiding the \troperef{rubber}{Rubber-Stamp trope}.

\paragraph{Quantify effect}
Variability between independent manual relevance labels and human-verified LLM labels can reveal the extent of black-box behavior. 


\paragraph{Guardrail} 
To mitigate this issue, complex labeling tasks should be broken into smaller steps, each with explicit reasoning guidelines \citep{mayfield2024evaluation,farzi2024exampp}. Both LLMs and human judges should articulate their reasoning at multiple stages, which makes decisions more interpretable and auditable. Stepwise reasoning, inspired by chain-of-thought prompting in GPT models, can increase transparency and robustness in evaluation.

\tropeheading{14}{Judge}{Predictable Secrets}{When human data can be guessed by an LLM.}{guess}

Many evaluation paradigms incorporate \emph{secrets}, information known only to human judges and withheld from the system, to prevent evaluation leakage. These include human-generated relevance labels, grading rubrics \citep{farzi2024exampp}, or nugget annotations \citep{lin2006different}. Such mechanisms are designed to guard against the negative effects of LLM-based evaluation.

However, these guardrails become ineffective when an LLM can reliably infer the secret. This introduces inadvertent circularity and undermines the purpose of human oversight \cite{deng2024investigating}. Predictable secrets typically signal that test points are too simplistic or follow an obvious pattern. 
Evaluation labels generated or structured by LLMs may exhibit consistent patterns that make them predictable and leakable. This allows systems to infer and exploit the evaluation signal, even in good-faith settings \citep{soboroff2025don,li2025preference}.

If an IR system can use an LLM to anticipate the secret and incorporate it into its output, it may achieve inflated scores, despite the apparent use of human judgment in the evaluation pipeline.

\paragraph{Quantify effect}
The guessability of a secret can be measured by having an LLM predict secrets directly or by computing its per-token likelihood. Downstream effects can be evaluated by replacing the manually created secret with the predicted secret and observing its impact on system rankings.

\paragraph{Guardrails}
When validity of the evaluation relies on secret information known only to human judges, it is essential to ensure that the secret is complex and varied enough to resist LLM inference. Designing tasks where secrets require true contextual understanding or subjective reasoning can help maintain this barrier.

%% file: case-studies.tex
\section{Case Studies}
\label{sec:usecases}

To demonstrate that our listed LLM-Evaluation tropes are in fact real issues, we
explore two case studies of LLM evaluation methods used in industry and which
guardrails were implemented to combat the risks.\footnote{Company names
anonymized for review.}

\subsection{\Canva{}}
\input{canva-case-study}

\subsection{\Valence{}} This case study examines AI-assisted enterprise
coaching, where sessions address complex workplace challenges. The system
integrates multiple large commercial LLMs with specialized dialogue components
for domain expertise, personalized memory, and user work profiles.

An LLM-based evaluator assesses dialogue quality at both conversational and turn
levels, incorporating Client Satisfaction (CSAT) scores and proprietary coaching
effectiveness measures. The evaluation methodology follows a rubric-based
framework, similar to \citet{lin2024interpretableusersatisfactionestimation} but
manually adapted to coaching tasks by subject matter experts. The evaluation
framework serves key purposes including:
\begin{itemize}
    \item Privacy-Preserving Evaluation: Assesses dialogue quality without
          exposing sensitive conversations to human reviewers.
    \item LLM-Based User Simulation: Tests alternative prompts, system
          configurations, and model components through synthetic interactions.
    \item LLM as an Autonomous Judge: Enables offline optimization of coaching
          effectiveness across key conversational dimensions.
\end{itemize}
A key challenge is \troperef{narcissism}{LLM Narcissism}, leading to inflated effectiveness estimates
compared to human assessments. To mitigate this, we use separate LLMs for
generation and scoring, calibrated against human-labeled datasets.

As our system evolves, enhancing LLM-based evaluation is critical for
scalability, privacy, and expert-level assessment quality. Two emerging
challenges stand out: 1) LLM as a Reward Model for RL Optimization, while
promising, risks reward hacking, where the model optimizes for evaluation
heuristics rather than genuine coaching effectiveness. 2) There is a shift to
real-time evaluation where evaluation moves from offline to online assessment at
the turn-level to enable active intervention when dialogue quality drops, but
could further amplify \troperef{circularity}{Circularity} biases potentially creating
negative feedback loops such as \troperef{concept-drift}{Self-training Collapse} and \troperef{synthetic}{Overfitting}.


%% file: canva-case-study.tex
In this case study (detailed in \citet{Cotterill2024}), the LLM-evaluator is
used in a known-item or re-finding task. It is particularly valuable in a
private or enterprise search environment, in which queries and documents are not
readily available, let alone viewable by system developers due to privacy
restrictions. Rather than providing relevance labels over a corpus of items,
instead the LLM is used to synthesize a known-item according to some desired
properties. Any number of additional items are  also generated, both ones that
are ``distant'' from the target item and ones that are ``near'' to the target
item. It generates one or more queries that represent a user trying to re-find
the target item. The characteristics of items and queries are derived from
anonymized aggregated statistics over the real user data, thereby grounding the
LLM-evaluator, avoiding \troperef{circularity}{Circularity trope}.  In this way, we generated a
conventional test collection, but with the properties that the relevance
judgments are known at inception, rather than requiring subsequent human
annotation.

The goals of this setup are three-fold:
\begin{itemize}
\item Eliminate privacy challenges allowing conventional eyes-on analysis and
debugging of search systems.
\item Directly construct retrieval and ranking challenges that match specific
areas for product improvement e.g., spell correction.
\item Ensure repeatability of evaluation, through archiving of generated test
collections. Changes to the search system are efficiently and deterministically
evaluated offline, accelerating rejection of bad improvements before testing
with people.
\end{itemize}

Although this offline evaluation was then succeeded by online interleaving and
A/B experiments, we demonstrated that, provided the improvements we observe in
the entirely synthetic LLM evaluation framework are directionally aligned to
these later-stage human-centered evaluations, we have no need to also involve
humans in the first stage.  
Both the known-item task (exactly one right answer) and involvement of humans at
later-stage evaluations de-risk \troperef{narcissism}{LLM Narcissism} and \troperef{circularity}{Circularity}; we had no LLM
involved in the search system either. 


%% file: demonstration.tex
\section{Demonstration of Circularity}
\label{sec:demonstration}

We simulate the effects of circularity on data from the TREC Retrieval-augmented Generation track (TREC RAG
2024) \cite{upadhyay2024large}. Given the recency systems used in this dataset, we reduce the risk of the \troperef{old-systems}{Old Systems trope}. Omitting very low performing systems, we consider the top 60 submitted results of TREC RAG 24 retrieval systems, called ``original'' systems in the following. 
We simulate the \troperef{eval_as_ranker}{LLM Eval as a Ranker} trope by re-ranking all original systems with the {\sc Umbrela} system, which we refer to as re-ranked systems for short. Specifically each
system run is re-ranked using the official qrels obtained via {\sc Umbrela}, with the {\sc
Umbrela} grade as the primary key and the original rank score as the  secondary
key. By TREC standards, the result would qualify as an automatic run. Our findings corroborate the discussion of
\citet{clarke2024llm} and show the effects of \troperef{circularity}{Circularity}.

\paragraph{LLM-evaluators are good re-rankers.}
We confirm that it is objectively beneficial for a system to incorporate
approaches from LLM-evaluators as suggested in prior
work~\cite{ma2023finetuning,ma2023zeroshot}. We simulate this by reranking the
outputs of all retrieval systems with {\sc Umbrela}. 
Comparing original and re-ranked systems on manual judgments in Figure~\ref{fig:rag24-llmranker}, we see that re-ranking with {\sc Umbrela}
indeed has a positive impact on the system performance. The (valid) conclusion
is that any system developer should be encouraged to adopt this approach. We expect more IR systems to adopt this approach in the near-future.

\begin{figure}[t]
    \centering
    \includegraphics[width=1\linewidth]{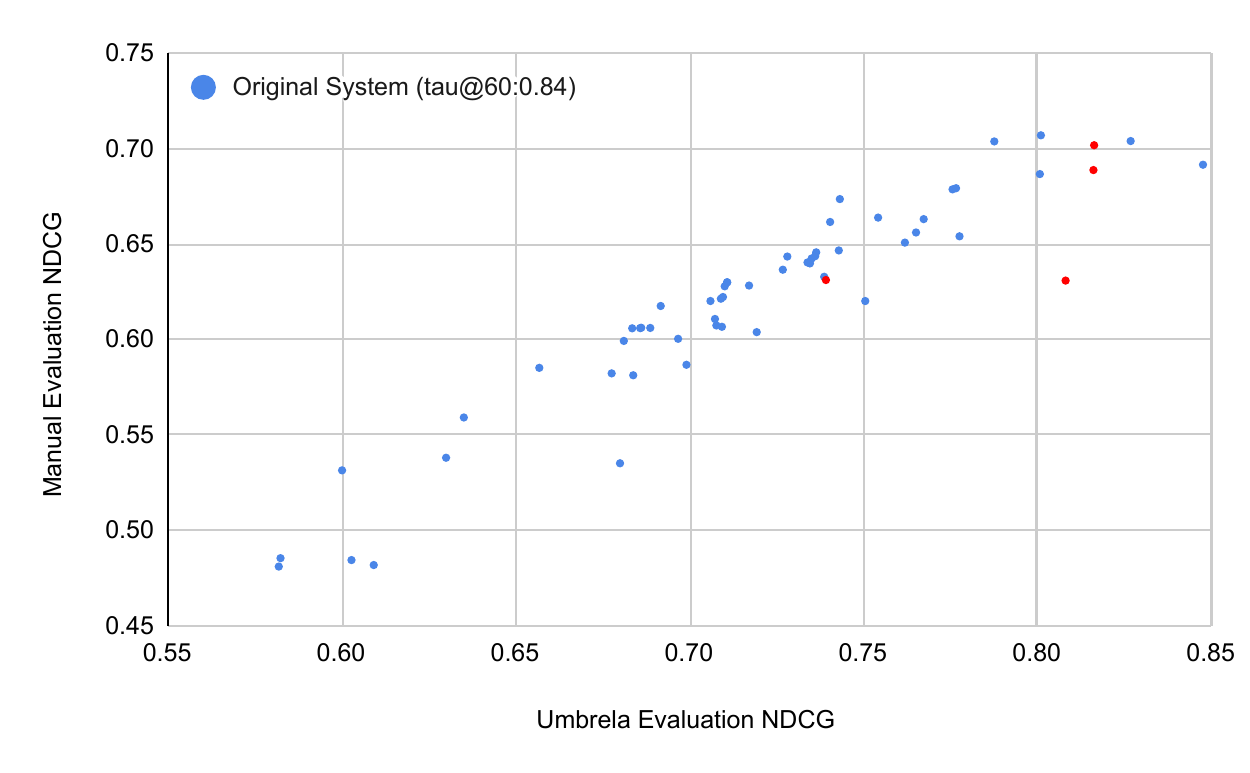}
    \caption{Reproduction of \citet{upadhyay2024umbrela}: On top 60 original TREC RAG 24 systems and data, the {\sc Umbrela} LLM evaluator correlates highly with manual assessors. Only few submitted retrieval systems included approaches from LLM evaluators.  Each
    system represents one dot. Red dots mark systems known to contain LLM evaluators \cite{clarke2024llm}.}
    \label{fig:rag24-evaluator}
\end{figure}

\begin{figure}[t]
    \centering
    \includegraphics[width=1\linewidth]{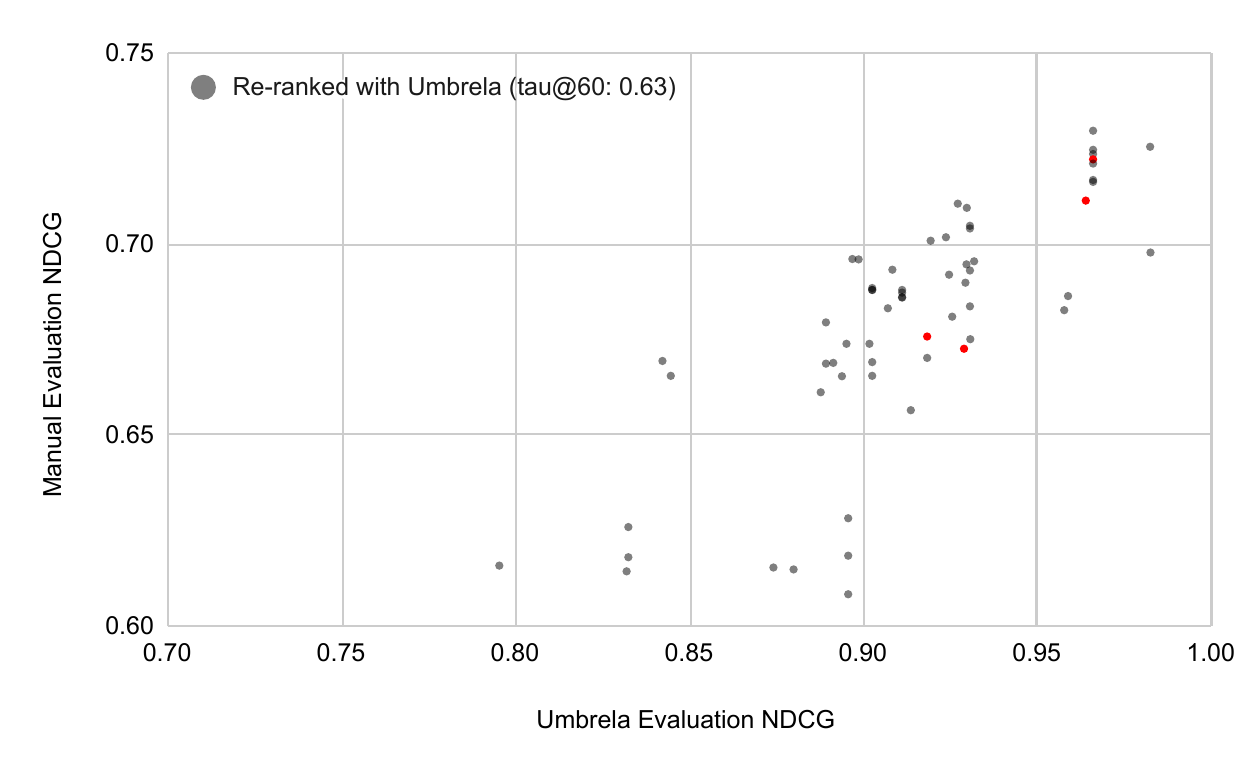}
    \caption{Demonstration of the effects of circularity when using {\sc Umbrela} as both evluator and ranker using TREC RAG 24 data. Each submitted retrieval
    system is first re-ranked with {\sc Umbrela}, then evaluated under NDCG with
    relevance labels from human judges and the {\sc Umbrela} evaluator. We see that
    especially among top ranked systems, the evaluation strategy no longer agrees
    with human judges on which system is better. Showing same top 60 (of 75) systems as in Figure \ref{fig:rag24-evaluator}.}
    \label{fig:rag24-llmranker-evaluator}
\end{figure}

\paragraph{LLM-evaluators are valid on systems that don't use LLM evaluators.}
On the original systems, track organizers \cite{upadhyay2024large} demonstrate with a Kendall's tau leaderboard correlation test that using {\sc Umbrela} as LLM-evaluator leads to very similar results as when using human judges. Using the top 60 systems in Figure \ref{fig:rag24-evaluator}, we confirm that this results in a relatively high Kendall's tau
measure of 0.84, which relates to about 8\% of discordant system pairs, i.e., a
pair of systems where the two evaluators disagree on which system is better. Overall this confirms the findings of \citet{upadhyay2024large}. We note one a few outliers, such as one system in Figure \ref{fig:rag24-evaluator} has a high LLM evaluation score (0.81) but a lower manual score (0.63). This systems includes an LLM evaluator as part of the system, further described in \citet{clarke2024llm}.

\paragraph{Circularity arises when LLM evaluators are used both by the system (for re-ranking) and for evaluation.}
We would obtain an invalid evaluation paradigm if we were to use the
{\sc Umbrela} evaluation metric  to evaluate these {\sc Umbrela} re-ranked
systems. This effect is demonstrated in Figure
\ref{fig:rag24-llmranker-evaluator}, where the {\sc Umbrela}-reranked system
runs are evaluated with NDCG under  {\sc Umbrela} relevance labels and manual
relevance judgments -- the latter measurement also being presented in Figure
\ref{fig:rag24-llmranker}.

We observe an increased disagreement between both evaluation systems, resulting
in a higher number of discordant pairs (18\% within the top 60).
This results in a much lower Kendall's tau metric of only 0.63. 

This statistic further decreases to 0.44 when only the top 20 systems are compared.  We note that for most research publications, it matters to demonstrate that all best systems are significantly outperformed by the proposed new system. Under this circular evaluation setup, such findings are not believable.

Additionally, we find that under the {\sc Umbrela} evaluation, twelve systems now
obtain an NDCG score above 0.95  which would suggest a near-perfect ranking quality that cannot be further improved. 
However, same systems only obtain manual NDCG scores between 0.68--0.72,
demonstrating the inflation of evaluation scores due to circularity.

\bigskip 

Taking manually created relevance labels as the gold standard, we conclude that
using the {\sc Umbrela} LLM evaluator on systems that internally (may) utilize
an {\sc Umbrela} re-ranking approach leads to an invalid experimental
evaluation.

\bigskip



\ld{more analysis:
\url{https://docs.google.com/spreadsheets/d/1EwQssYQtrnaADgd8wffCZdo0BZftAnVt1yQeJKtXv2c/edit?usp=sharing}}

%% file: required-experiments.tex
\section{Suggested Experimentation Infrastructure}

\label{sec:coopetition}

Comparing LLM-based and traditional evaluation metrics on the same IR systems is
essential to quantify biases and inconsistencies.

We propose a TREC-style collaborative competition, a so-called ``Coopetition'',\footnote{Coopetition refers to cooperative
    competition, where multiple research groups collaboratively compete to
    identify the most effective LLM-based evaluator, grounded in manual
    assessments, for use in academic research.} structured around a shared task
    with a predefined set of topics. Participants would submit contributions in
    three categories:
\begin{description}[style=unboxed,leftmargin=0cm]
    \item [1. IR Systems] that attempt to solve the task using retrieval-based,
          generative, or mixed-modality approaches.
    \item [2. LLM Evaluators] that assess system outputs, either by ranking
          systems or generating relevance judgments.
    \item [3. Content Modification Strategies] to deliberately alter documents
    with the goal of testing system and evaluator robustness.
\end{description}

The expected outcome is a test collection with human-verified relevance labels,
along with one or more LLM-based evaluators that demonstrate strong performance.
The inclusion of modified content introduces an adversarial component,
stress-testing both retrieval systems and evaluators. This setup encourages the
development of more robust methods while helping to identify potential risks and
vulnerabilities in a controlled environment. Ideally, the top-performing LLM
evaluator(s) should have an open-source implementation, ensuring transparency
and enabling their use for addressing future gaps in judgment coverage. Beyond
identifying the best-performing systems, this initiative should serve as an
ongoing benchmarking effort, similar to leaderboard-style evaluations. Rather
than emphasizing marginal performance gains, the goal is to build a diverse set
of baseline systems and evaluators for meaningful comparisons.

We envision the Coopetition as an annual effort, introducing new tasks, topics,
modified content, and fresh human relevance judgments in each iteration. If a
superior LLM evaluator emerges, it will replace the previous best, updating the
test collection accordingly. Between iterations, researchers would be encouraged
to use the best available LLM evaluator for offline experiments, method
development, and publications. A key aspect of this initiative is to obtain a
resilient evaluation even when content is modified in adversarial ways. It will
also help identify what makes an LLM-based evaluation system effective, where it
fails, and how it can be improved; followed by defining standardized guard rails
for validating experiments.









%% file: conclusion.tex
\section{Conclusions}

With this paper we aim to codify the best practices for ensuring that LLM-based
evaluation remains a valid experimentation approach for IR research. Maintaining
scientific rigor requires identifying and recognizing the risks associated with
synthetic training data  and LLM-based ranking, while ensuring they
are cross-validated with human-verified benchmarks. Automatic
evaluation methodologies must be adopted cautiously, treating them as
validation tools rather than definitive measures of system performance.

To address these challenges, we propose a new form of TREC-style Coopetition
which annually identifies the best LLM-evaluation approaches measuring
state-of-the-art IR systems on fresh test collections. This would ensure we are
using (1) the best technology, and (2) continuously confirming LLM-evaluation
validity with manual judgments.

\balance
\paragraph{Can I use LLM-judgments in my next conference paper?}

We conclude that LLM-based judgments can be used, but only under certain conditions that ensure evaluation validity:

\begin{itemize}
    \item If the LLM-based evaluation metrics have been recently validated
    against real user judgments and are supported by diverse complementary
    metrics to prevent overfitting or bias.

    \item If the evaluation setup ensures that LLM evaluators do not
    inadvertently influence system development, with potential risks such as
    test signal leakage and circularity being demonstrably mitigated.

    \item If potential biases and evaluation tropes associated with the LLM
    evaluator are thoroughly discussed, quantified, and addressed through
    effective guardrails.
\end{itemize}

Applying this framework ensures that LLM-based evaluations remain trustworthy,
reproducible, and scientifically grounded -- reducing the danger of developing
and deploying systems that are of limited utility for (human) users.

%% file: main.bbl

\begin{thebibliography}{84}


\ifx \showCODEN    \undefined \def \showCODEN     #1{\unskip}     \fi
\ifx \showISBNx    \undefined \def \showISBNx     #1{\unskip}     \fi
\ifx \showISBNxiii \undefined \def \showISBNxiii  #1{\unskip}     \fi
\ifx \showISSN     \undefined \def \showISSN      #1{\unskip}     \fi
\ifx \showLCCN     \undefined \def \showLCCN      #1{\unskip}     \fi
\ifx \shownote     \undefined \def \shownote      #1{#1}          \fi
\ifx \showarticletitle \undefined \def \showarticletitle #1{#1}   \fi
\ifx \showURL      \undefined \def \showURL       {\relax}        \fi
\providecommand\bibfield[2]{#2}
\providecommand\bibinfo[2]{#2}
\providecommand\natexlab[1]{#1}
\providecommand\showeprint[2][]{arXiv:#2}

\bibitem[Abbasiantaeb et~al\mbox{.}(2024)]%
        {abbasiantaeb2024can}
\bibfield{author}{\bibinfo{person}{Zahra Abbasiantaeb}, \bibinfo{person}{Chuan
  Meng}, \bibinfo{person}{Leif Azzopardi}, {and} \bibinfo{person}{Mohammad
  Aliannejadi}.} \bibinfo{year}{2024}\natexlab{}.
\newblock \showarticletitle{Can We Use Large Language Models to Fill Relevance
  Judgment Holes?}. In \bibinfo{booktitle}{\emph{EMTCIR ’24: The First
  Workshop on Evaluation Methodologies, Testbeds and Community for Information
  Access Research}}.
\newblock


\bibitem[Alaofi et~al\mbox{.}(2024a)]%
        {alaofi2024generative}
\bibfield{author}{\bibinfo{person}{Marwah Alaofi}, \bibinfo{person}{Negar
  Arabzadeh}, \bibinfo{person}{Charles~LA Clarke}, {and} \bibinfo{person}{Mark
  Sanderson}.} \bibinfo{year}{2024}\natexlab{a}.
\newblock \showarticletitle{Generative information retrieval evaluation}.
\newblock In \bibinfo{booktitle}{\emph{Information Access in the Era of
  Generative AI}}. \bibinfo{publisher}{Springer}, \bibinfo{pages}{135--159}.
\newblock


\bibitem[Alaofi et~al\mbox{.}(2024b)]%
        {Alaofi2024}
\bibfield{author}{\bibinfo{person}{Marwah Alaofi}, \bibinfo{person}{Paul
  Thomas}, \bibinfo{person}{Falk Scholer}, {and} \bibinfo{person}{Mark
  Sanderson}.} \bibinfo{year}{2024}\natexlab{b}.
\newblock \showarticletitle{LLMs can be Fooled into Labelling a Document as
  Relevant: best caf\'{e} near me; this paper is perfectly relevant}. In
  \bibinfo{booktitle}{\emph{Proceedings of the 2024 Annual International ACM
  SIGIR Conference on Research and Development in Information Retrieval in the
  Asia Pacific Region}} \emph{(\bibinfo{series}{SIGIR-AP 2024})}.
  \bibinfo{pages}{32--41}.
\newblock


\bibitem[Anonymous(2024)]%
        {Cotterill2024}
\bibfield{author}{\bibinfo{person}{Author Anonymous}.}
  \bibinfo{year}{2024}\natexlab{}.
\newblock \bibinfo{title}{{How to improve search without looking at queries or
  results}}.
\newblock
\urldef\tempurl%
\url{anonymized/for/review}
\showURL{%
\tempurl}


\bibitem[Arabzadeh and Clarke(2025)]%
        {arabzadeh2025human}
\bibfield{author}{\bibinfo{person}{Negar Arabzadeh} {and}
  \bibinfo{person}{Charles~LA Clarke}.} \bibinfo{year}{2025}\natexlab{}.
\newblock \showarticletitle{A Human-AI Comparative Analysis of Prompt
  Sensitivity in LLM-Based Relevance Judgment}.
\newblock \bibinfo{journal}{\emph{arXiv preprint arXiv:2504.12408}}
  (\bibinfo{year}{2025}).
\newblock


\bibitem[Asch(1958)]%
        {asch1958effects}
\bibfield{author}{\bibinfo{person}{Solomon Asch}.}
  \bibinfo{year}{1958}\natexlab{}.
\newblock \showarticletitle{Effects of group pressure on the modification and
  distortion}.
\newblock \bibinfo{journal}{\emph{Readings in social psychology. New York:
  Holt, Rinehart and Winston}} (\bibinfo{year}{1958}).
\newblock


\bibitem[Authors(2023)]%
        {distrust2023}
\bibfield{author}{\bibinfo{person}{Authors}.} \bibinfo{year}{2023}\natexlab{}.
\newblock \showarticletitle{The Importance of Distrust in AI}.
\newblock \bibinfo{journal}{\emph{arXiv preprint arXiv:2307.13601}}
  (\bibinfo{year}{2023}).
\newblock
\urldef\tempurl%
\url{https://arxiv.org/abs/2307.13601}
\showURL{%
\tempurl}


\bibitem[Balog et~al\mbox{.}(2025)]%
        {balog2025rankers}
\bibfield{author}{\bibinfo{person}{Krisztian Balog}, \bibinfo{person}{Donald
  Metzler}, {and} \bibinfo{person}{Zhen Qin}.} \bibinfo{year}{2025}\natexlab{}.
\newblock \showarticletitle{Rankers, Judges, and Assistants: Towards
  Understanding the Interplay of LLMs in Information Retrieval Evaluation}.
\newblock \bibinfo{journal}{\emph{arXiv preprint arXiv:2503.19092}}
  (\bibinfo{year}{2025}).
\newblock


\bibitem[Bardas et~al\mbox{.}(2025)]%
        {bardas2025}
\bibfield{author}{\bibinfo{person}{Niv Bardas}, \bibinfo{person}{Tommy Mordo},
  \bibinfo{person}{Oren Kurland}, \bibinfo{person}{Moshe Tennenholtz}, {and}
  \bibinfo{person}{Gal Zur}.} \bibinfo{year}{2025}\natexlab{}.
\newblock \showarticletitle{Prompt-Based Document Modifications in Ranking
  Competitions}.
\newblock \bibinfo{journal}{\emph{arXiv preprint arXiv:2502.07315}}
  (\bibinfo{year}{2025}).
\newblock


\bibitem[Ben~Basat et~al\mbox{.}(2015)]%
        {basat2015}
\bibfield{author}{\bibinfo{person}{Ran Ben~Basat}, \bibinfo{person}{Moshe
  Tennenholtz}, {and} \bibinfo{person}{Oren Kurland}.}
  \bibinfo{year}{2015}\natexlab{}.
\newblock \showarticletitle{The Probability Ranking Principle is Not Optimal in
  Adversarial Retrieval Settings}. In \bibinfo{booktitle}{\emph{Proceedings of
  the 2015 International Conference on The Theory of Information Retrieval}}
  \emph{(\bibinfo{series}{ICTIR '15})}. \bibinfo{pages}{51–60}.
\newblock


\bibitem[Bhatt and Diaz(2024)]%
        {bhatt2024extrinsic}
\bibfield{author}{\bibinfo{person}{Shaily Bhatt} {and}
  \bibinfo{person}{Fernando Diaz}.} \bibinfo{year}{2024}\natexlab{}.
\newblock \showarticletitle{Extrinsic Evaluation of Cultural Competence in
  Large Language Models}. In \bibinfo{booktitle}{\emph{Findings of the
  Association for Computational Linguistics: EMNLP 2024}}.
  \bibinfo{pages}{12345--12356}.
\newblock


\bibitem[Bordt et~al\mbox{.}(2024)]%
        {bordt2024elephants}
\bibfield{author}{\bibinfo{person}{Sebastian Bordt}, \bibinfo{person}{Harsha
  Nori}, \bibinfo{person}{Vanessa Rodrigues}, \bibinfo{person}{Besmira Nushi},
  {and} \bibinfo{person}{Rich Caruana}.} \bibinfo{year}{2024}\natexlab{}.
\newblock \showarticletitle{Elephants Never Forget: Memorization and Learning
  of Tabular Data in Large Language Models}.
\newblock \bibinfo{journal}{\emph{arXiv preprint arXiv:2404.06209}}
  (\bibinfo{year}{2024}).
\newblock


\bibitem[Chen et~al\mbox{.}(2023)]%
        {chen2023chatgpt}
\bibfield{author}{\bibinfo{person}{Lingjiao Chen}, \bibinfo{person}{Matei
  Zaharia}, {and} \bibinfo{person}{James Zou}.}
  \bibinfo{year}{2023}\natexlab{}.
\newblock \showarticletitle{How is ChatGPT's behavior changing over time?}
\newblock \bibinfo{journal}{\emph{arXiv preprint arXiv:2307.09009}}
  (\bibinfo{year}{2023}).
\newblock


\bibitem[Clarke et~al\mbox{.}(2016)]%
        {clarke2016assessing}
\bibfield{author}{\bibinfo{person}{Charles~LA Clarke}, \bibinfo{person}{J~Shane
  Culpepper}, {and} \bibinfo{person}{Alistair Moffat}.}
  \bibinfo{year}{2016}\natexlab{}.
\newblock \showarticletitle{Assessing efficiency--effectiveness tradeoffs in
  multi-stage retrieval systems without using relevance judgments}.
\newblock \bibinfo{journal}{\emph{Information Retrieval Journal}}
  \bibinfo{volume}{19} (\bibinfo{year}{2016}), \bibinfo{pages}{351--377}.
\newblock


\bibitem[Clarke and Dietz(2024)]%
        {clarke2024llm}
\bibfield{author}{\bibinfo{person}{Charles~LA Clarke} {and}
  \bibinfo{person}{Laura Dietz}.} \bibinfo{year}{2024}\natexlab{}.
\newblock \showarticletitle{LLM-based relevance assessment still can't replace
  human relevance assessment}.
\newblock \bibinfo{journal}{\emph{arXiv preprint arXiv:2412.17156}}
  (\bibinfo{year}{2024}).
\newblock


\bibitem[Claypoole et~al\mbox{.}(2019)]%
        {claypoole2019effects}
\bibfield{author}{\bibinfo{person}{Victoria~L Claypoole},
  \bibinfo{person}{Daryn~A Dever}, \bibinfo{person}{Kody~L Denues}, {and}
  \bibinfo{person}{James~L Szalma}.} \bibinfo{year}{2019}\natexlab{}.
\newblock \showarticletitle{The effects of event rate on a cognitive vigilance
  task}.
\newblock \bibinfo{journal}{\emph{Human factors}} \bibinfo{volume}{61},
  \bibinfo{number}{3} (\bibinfo{year}{2019}), \bibinfo{pages}{440--450}.
\newblock


\bibitem[Cleverdon(1967)]%
        {cleverdon1967}
\bibfield{author}{\bibinfo{person}{Cyril Cleverdon}.}
  \bibinfo{year}{1967}\natexlab{}.
\newblock \showarticletitle{The Cranfield Tests on Index Language Devices}.
\newblock \bibinfo{journal}{\emph{Aslib Proceedings}} \bibinfo{volume}{19},
  \bibinfo{number}{6} (\bibinfo{year}{1967}), \bibinfo{pages}{173--194}.
\newblock


\bibitem[Cormack and Grossman(2015)]%
        {cormack2015autonomy}
\bibfield{author}{\bibinfo{person}{Gordon~V Cormack} {and}
  \bibinfo{person}{Maura~R Grossman}.} \bibinfo{year}{2015}\natexlab{}.
\newblock \showarticletitle{Autonomy and reliability of continuous active
  learning for technology-assisted review}.
\newblock \bibinfo{journal}{\emph{arXiv preprint arXiv:1504.06868}}
  (\bibinfo{year}{2015}).
\newblock


\bibitem[Cormack and Grossman(2016)]%
        {cormack2016scalability}
\bibfield{author}{\bibinfo{person}{Gordon~V Cormack} {and}
  \bibinfo{person}{Maura~R Grossman}.} \bibinfo{year}{2016}\natexlab{}.
\newblock \showarticletitle{Scalability of continuous active learning for
  reliable high-recall text classification}. In
  \bibinfo{booktitle}{\emph{Proceedings of the 25th ACM international on
  conference on information and knowledge management}}.
  \bibinfo{pages}{1039--1048}.
\newblock


\bibitem[Deng et~al\mbox{.}(2024)]%
        {deng2024investigating}
\bibfield{author}{\bibinfo{person}{Chunyuan Deng}, \bibinfo{person}{Yilun
  Zhao}, \bibinfo{person}{Xiangru Tang}, \bibinfo{person}{Mark Gerstein}, {and}
  \bibinfo{person}{Arman Cohan}.} \bibinfo{year}{2024}\natexlab{}.
\newblock \showarticletitle{Investigating Data Contamination in Modern
  Benchmarks for Large Language Models}. In
  \bibinfo{booktitle}{\emph{Proceedings of the 2024 Conference of the North
  American Chapter of the Association for Computational Linguistics: Human
  Language Technologies (Volume 1: Long Papers)}}. \bibinfo{pages}{8698--8711}.
\newblock


\bibitem[Dohmatob et~al\mbox{.}(2024)]%
        {dohmatob2024demystified}
\bibfield{author}{\bibinfo{person}{Elvis Dohmatob}, \bibinfo{person}{Yunzhen
  Feng}, {and} \bibinfo{person}{Julia Kempe}.} \bibinfo{year}{2024}\natexlab{}.
\newblock \showarticletitle{Model Collapse Demystified: The Case of
  Regression}.
\newblock \bibinfo{journal}{\emph{arXiv preprint arXiv:2402.07712}}
  (\bibinfo{year}{2024}).
\newblock


\bibitem[Faggioli et~al\mbox{.}(2023)]%
        {faggioli2023perspectives}
\bibfield{author}{\bibinfo{person}{Guglielmo Faggioli}, \bibinfo{person}{Laura
  Dietz}, \bibinfo{person}{Charles~LA Clarke}, \bibinfo{person}{Gianluca
  Demartini}, \bibinfo{person}{Matthias Hagen}, \bibinfo{person}{Claudia
  Hauff}, \bibinfo{person}{Noriko Kando}, \bibinfo{person}{Evangelos Kanoulas},
  \bibinfo{person}{Martin Potthast}, \bibinfo{person}{Benno Stein},
  {et~al\mbox{.}}} \bibinfo{year}{2023}\natexlab{}.
\newblock \showarticletitle{Perspectives on large language models for relevance
  judgment}. In \bibinfo{booktitle}{\emph{Proceedings of the 2023 ACM SIGIR
  International Conference on Theory of Information Retrieval}}.
  \bibinfo{pages}{39--50}.
\newblock


\bibitem[Farzi and Dietz(2024)]%
        {farzi2024exampp}
\bibfield{author}{\bibinfo{person}{Naghmeh Farzi} {and} \bibinfo{person}{Laura
  Dietz}.} \bibinfo{year}{2024}\natexlab{}.
\newblock \showarticletitle{Exam++: Llm-based answerability metrics for ir
  evaluation}. In \bibinfo{booktitle}{\emph{Proceedings of LLM4Eval: The First
  Workshop on Large Language Models for Evaluation in Information Retrieval}}.
\newblock


\bibitem[Fok and Weld(2024)]%
        {fok2024search}
\bibfield{author}{\bibinfo{person}{Raymond Fok} {and} \bibinfo{person}{Daniel~S
  Weld}.} \bibinfo{year}{2024}\natexlab{}.
\newblock \showarticletitle{In search of verifiability: Explanations rarely
  enable complementary performance in AI-advised decision making}.
\newblock \bibinfo{journal}{\emph{AI Magazine}} \bibinfo{volume}{45},
  \bibinfo{number}{3} (\bibinfo{year}{2024}), \bibinfo{pages}{317--332}.
\newblock


\bibitem[Fr\"{o}be et~al\mbox{.}(2023)]%
        {Frobe2023}
\bibfield{author}{\bibinfo{person}{Maik Fr\"{o}be}, \bibinfo{person}{Lukas
  Gienapp}, \bibinfo{person}{Martin Potthast}, {and} \bibinfo{person}{Matthias
  Hagen}.} \bibinfo{year}{2023}\natexlab{}.
\newblock \showarticletitle{Bootstrapped nDCG Estimation in the Presence of
  Unjudged Documents}. In \bibinfo{booktitle}{\emph{Advances in Information
  Retrieval: 45th European Conference on Information Retrieval, ECIR 2023,
  Dublin, Ireland, April 2–6, 2023, Proceedings, Part I}}.
  \bibinfo{pages}{313–329}.
\newblock


\bibitem[Gao et~al\mbox{.}(2024)]%
        {gao2024llmenhanced}
\bibfield{author}{\bibinfo{person}{Jingtong Gao}, \bibinfo{person}{Bo Chen},
  \bibinfo{person}{Xiangyu Zhao}, \bibinfo{person}{Weiwen Liu},
  \bibinfo{person}{Xiangyang Li}, \bibinfo{person}{Yichao Wang},
  \bibinfo{person}{Zijian Zhang}, \bibinfo{person}{Wanyu Wang},
  \bibinfo{person}{Yuyang Ye}, \bibinfo{person}{Shanru Lin},
  \bibinfo{person}{Huifeng Guo}, {and} \bibinfo{person}{Ruiming Tang}.}
  \bibinfo{year}{2024}\natexlab{}.
\newblock \showarticletitle{LLM-enhanced Reranking in Recommender Systems}.
\newblock \bibinfo{journal}{\emph{arXiv preprint arXiv:2406.12433}}
  (\bibinfo{year}{2024}).
\newblock


\bibitem[Gerstgrasser et~al\mbox{.}(2024)]%
        {gerstgrasser2024model}
\bibfield{author}{\bibinfo{person}{Matthias Gerstgrasser},
  \bibinfo{person}{Rylan Schaeffer}, \bibinfo{person}{Apratim Dey},
  \bibinfo{person}{Rafael Rafailov}, \bibinfo{person}{Henry Sleight},
  \bibinfo{person}{John Hughes}, \bibinfo{person}{Tomasz Korbak},
  \bibinfo{person}{Rajashree Agrawal}, \bibinfo{person}{Dhruv Pai},
  \bibinfo{person}{Andrey Gromov}, \bibinfo{person}{Daniel~A. Roberts},
  \bibinfo{person}{Diyi Yang}, \bibinfo{person}{David~L. Donoho}, {and}
  \bibinfo{person}{Sanmi Koyejo}.} \bibinfo{year}{2024}\natexlab{}.
\newblock \showarticletitle{Is Model Collapse Inevitable? Breaking the Curse of
  Recursion by Accumulating Real and Synthetic Data}.
\newblock \bibinfo{journal}{\emph{arXiv preprint arXiv:2404.01413}}
  (\bibinfo{year}{2024}).
\newblock


\bibitem[Goodhart(1975)]%
        {goodhart1975problems}
\bibfield{author}{\bibinfo{person}{Charles Goodhart}.}
  \bibinfo{year}{1975}\natexlab{}.
\newblock \showarticletitle{Problems of monetary management: the UK experience
  in papers in monetary economics}.
\newblock \bibinfo{journal}{\emph{Monetary Economics}}  \bibinfo{volume}{1}
  (\bibinfo{year}{1975}).
\newblock


\bibitem[Griffin et~al\mbox{.}(2023)]%
        {griffin2023susceptibility}
\bibfield{author}{\bibinfo{person}{Lewis~D Griffin}, \bibinfo{person}{Bennett
  Kleinberg}, \bibinfo{person}{Maximilian Mozes}, \bibinfo{person}{Kimberly~T
  Mai}, \bibinfo{person}{Maria Vau}, \bibinfo{person}{Matthew Caldwell}, {and}
  \bibinfo{person}{Augustine Marvor-Parker}.} \bibinfo{year}{2023}\natexlab{}.
\newblock \showarticletitle{Susceptibility to influence of large language
  models}.
\newblock \bibinfo{journal}{\emph{arXiv preprint arXiv:2303.06074}}
  (\bibinfo{year}{2023}).
\newblock


\bibitem[Ji et~al\mbox{.}(2024)]%
        {ji2024detecting}
\bibfield{author}{\bibinfo{person}{Jiazhou Ji}, \bibinfo{person}{Ruizhe Li},
  \bibinfo{person}{Shujun Li}, \bibinfo{person}{Jie Guo},
  \bibinfo{person}{Weidong Qiu}, \bibinfo{person}{Zheng Huang},
  \bibinfo{person}{Chiyu Chen}, \bibinfo{person}{Xiaoyu Jiang}, {and}
  \bibinfo{person}{Xinru Lu}.} \bibinfo{year}{2024}\natexlab{}.
\newblock \showarticletitle{Detecting Machine-Generated Texts: Not Just "AI vs
  Humans" and Explainability is Complicated}.
\newblock \bibinfo{journal}{\emph{arXiv preprint arXiv:2406.18259}}
  (\bibinfo{year}{2024}).
\newblock


\bibitem[Kr{\"u}gel et~al\mbox{.}(2021)]%
        {krugel2021zombies}
\bibfield{author}{\bibinfo{person}{Sebastian Kr{\"u}gel},
  \bibinfo{person}{Andreas Ostermaier}, {and} \bibinfo{person}{Matthias Uhl}.}
  \bibinfo{year}{2021}\natexlab{}.
\newblock \showarticletitle{Zombies in the Loop? Humans Trust Untrustworthy
  AI-Advisors for Ethical Decisions}.
\newblock \bibinfo{journal}{\emph{arXiv preprint arXiv:2106.16122}}
  (\bibinfo{year}{2021}).
\newblock


\bibitem[Kurland and Tennenholtz(2022)]%
        {kurland2022}
\bibfield{author}{\bibinfo{person}{Oren Kurland} {and} \bibinfo{person}{Moshe
  Tennenholtz}.} \bibinfo{year}{2022}\natexlab{}.
\newblock \showarticletitle{Competitive Search}. In
  \bibinfo{booktitle}{\emph{Proceedings of the 45th International ACM SIGIR
  Conference on Research and Development in Information Retrieval}}
  \emph{(\bibinfo{series}{SIGIR '22})}. \bibinfo{pages}{2838–2849}.
\newblock


\bibitem[Li et~al\mbox{.}(2025a)]%
        {li2025llm}
\bibfield{author}{\bibinfo{person}{Ang Li}, \bibinfo{person}{Haozhe Chen},
  \bibinfo{person}{Hongseok Namkoong}, {and} \bibinfo{person}{Tianyi Peng}.}
  \bibinfo{year}{2025}\natexlab{a}.
\newblock \showarticletitle{LLM Generated Persona is a Promise with a Catch}.
\newblock \bibinfo{journal}{\emph{arXiv preprint arXiv:2503.16527}}
  (\bibinfo{year}{2025}).
\newblock


\bibitem[Li et~al\mbox{.}(2025b)]%
        {li2025preference}
\bibfield{author}{\bibinfo{person}{Dawei Li}, \bibinfo{person}{Renliang Sun},
  \bibinfo{person}{Yue Huang}, \bibinfo{person}{Ming Zhong},
  \bibinfo{person}{Bohan Jiang}, \bibinfo{person}{Jiawei Han},
  \bibinfo{person}{Xiangliang Zhang}, \bibinfo{person}{Wei Wang}, {and}
  \bibinfo{person}{Huan Liu}.} \bibinfo{year}{2025}\natexlab{b}.
\newblock \showarticletitle{Preference Leakage: A Contamination Problem in
  LLM-as-a-judge}.
\newblock \bibinfo{journal}{\emph{arXiv preprint arXiv:2502.01534}}
  (\bibinfo{year}{2025}).
\newblock


\bibitem[Liao and Vaughan(2023)]%
        {liao2023ai}
\bibfield{author}{\bibinfo{person}{Q.~Vera Liao} {and}
  \bibinfo{person}{Jennifer~Wortman Vaughan}.} \bibinfo{year}{2023}\natexlab{}.
\newblock \showarticletitle{AI Transparency in the Age of LLMs: A
  Human-Centered Research Roadmap}.
\newblock \bibinfo{journal}{\emph{arXiv preprint arXiv:2306.01941}}
  (\bibinfo{year}{2023}).
\newblock


\bibitem[Lin and Demner-Fushman(2007)]%
        {lin2006different}
\bibfield{author}{\bibinfo{person}{Jimmy Lin} {and} \bibinfo{person}{Dina
  Demner-Fushman}.} \bibinfo{year}{2007}\natexlab{}.
\newblock \showarticletitle{Different structures for evaluating answers to
  complex questions: Pyramids won't topple, and neither will human assessors}.
  In \bibinfo{booktitle}{\emph{Proceedings of the 45th Annual Meeting of the
  Association of Computational Linguistics}}. \bibinfo{pages}{561--568}.
\newblock


\bibitem[Lin et~al\mbox{.}(2024)]%
        {lin2024interpretableusersatisfactionestimation}
\bibfield{author}{\bibinfo{person}{Ying-Chun Lin}, \bibinfo{person}{Jennifer
  Neville}, \bibinfo{person}{Jack Stokes}, \bibinfo{person}{Longqi Yang},
  \bibinfo{person}{Tara Safavi}, \bibinfo{person}{Mengting Wan},
  \bibinfo{person}{Scott Counts}, \bibinfo{person}{Siddharth Suri},
  \bibinfo{person}{Reid Andersen}, \bibinfo{person}{Xiaofeng Xu},
  \bibinfo{person}{Deepak Gupta}, \bibinfo{person}{Sujay~Kumar Jauhar},
  \bibinfo{person}{Xia Song}, \bibinfo{person}{Georg Buscher},
  \bibinfo{person}{Saurabh Tiwary}, \bibinfo{person}{Brent Hecht}, {and}
  \bibinfo{person}{Jaime Teevan}.} \bibinfo{year}{2024}\natexlab{}.
\newblock \showarticletitle{Interpretable User Satisfaction Estimation for
  Conversational Systems with Large Language Models}. In
  \bibinfo{booktitle}{\emph{Proceedings of the 62nd Annual Meeting of the
  Association for Computational Linguistics}}. \bibinfo{pages}{11100--11115}.
\newblock


\bibitem[Liu et~al\mbox{.}(2023)]%
        {liu-etal-2023-g}
\bibfield{author}{\bibinfo{person}{Yang Liu}, \bibinfo{person}{Dan Iter},
  \bibinfo{person}{Yichong Xu}, \bibinfo{person}{Shuohang Wang},
  \bibinfo{person}{Ruochen Xu}, {and} \bibinfo{person}{Chenguang Zhu}.}
  \bibinfo{year}{2023}\natexlab{}.
\newblock \showarticletitle{{G}-Eval: {NLG} Evaluation using Gpt-4 with Better
  Human Alignment}. In \bibinfo{booktitle}{\emph{Proceedings of the 2023
  Conference on Empirical Methods in Natural Language Processing}}.
  \bibinfo{pages}{2511--2522}.
\newblock
\href{https://doi.org/10.18653/v1/2023.emnlp-main.153}{doi:\nolinkurl{10.18653/v1/2023.emnlp-main.153}}


\bibitem[Liu et~al\mbox{.}(2024)]%
        {liu2024llms}
\bibfield{author}{\bibinfo{person}{Yiqi Liu}, \bibinfo{person}{Nafise~Sadat
  Moosavi}, {and} \bibinfo{person}{Chenghua Lin}.}
  \bibinfo{year}{2024}\natexlab{}.
\newblock \showarticletitle{LLMs as Narcissistic Evaluators: When Ego Inflates
  Evaluation Scores}. In \bibinfo{booktitle}{\emph{Findings of the Association
  for Computational Linguistics ACL 2024}}. \bibinfo{pages}{12688--12701}.
\newblock


\bibitem[Lu et~al\mbox{.}(2017)]%
        {Lu2017}
\bibfield{author}{\bibinfo{person}{Xiaolu Lu}, \bibinfo{person}{Alistair
  Moffat}, {and} \bibinfo{person}{J~Shane Culpepper}.}
  \bibinfo{year}{2017}\natexlab{}.
\newblock \showarticletitle{Can Deep Effectiveness Metrics Be Evaluated Using
  Shallow Judgment Pools?}. In \bibinfo{booktitle}{\emph{Proceedings of the
  40th {International} {ACM} {SIGIR} {Conference} on {Research} and
  {Development} in {Information} {Retrieval}}}. \bibinfo{pages}{35--44}.
\newblock


\bibitem[Ma et~al\mbox{.}(2023a)]%
        {ma2023finetuning}
\bibfield{author}{\bibinfo{person}{Xueguang Ma}, \bibinfo{person}{Liang Wang},
  \bibinfo{person}{Nan Yang}, \bibinfo{person}{Furu Wei}, {and}
  \bibinfo{person}{Jimmy Lin}.} \bibinfo{year}{2023}\natexlab{a}.
\newblock \showarticletitle{Fine-Tuning LLaMA for Multi-Stage Text Retrieval}.
\newblock \bibinfo{journal}{\emph{arXiv preprint arXiv:2310.08319}}
  (\bibinfo{year}{2023}).
\newblock


\bibitem[Ma et~al\mbox{.}(2023b)]%
        {ma2023zeroshot}
\bibfield{author}{\bibinfo{person}{Xueguang Ma}, \bibinfo{person}{Xinyu Zhang},
  \bibinfo{person}{Ronak Pradeep}, {and} \bibinfo{person}{Jimmy Lin}.}
  \bibinfo{year}{2023}\natexlab{b}.
\newblock \showarticletitle{Zero-Shot Listwise Document Reranking with a Large
  Language Model}.
\newblock \bibinfo{journal}{\emph{arXiv preprint arXiv:2305.02156}}
  (\bibinfo{year}{2023}).
\newblock


\bibitem[Mayfield et~al\mbox{.}(2024)]%
        {mayfield2024evaluation}
\bibfield{author}{\bibinfo{person}{James Mayfield}, \bibinfo{person}{Eugene
  Yang}, \bibinfo{person}{Dawn Lawrie}, \bibinfo{person}{Sean MacAvaney},
  \bibinfo{person}{Paul McNamee}, \bibinfo{person}{Douglas~W Oard},
  \bibinfo{person}{Luca Soldaini}, \bibinfo{person}{Ian Soboroff},
  \bibinfo{person}{Orion Weller}, \bibinfo{person}{Efsun Kayi},
  {et~al\mbox{.}}} \bibinfo{year}{2024}\natexlab{}.
\newblock \showarticletitle{On the evaluation of machine-generated reports}. In
  \bibinfo{booktitle}{\emph{Proceedings of the 47th International ACM SIGIR
  Conference on Research and Development in Information Retrieval}}.
  \bibinfo{pages}{1904--1915}.
\newblock


\bibitem[Mehri and Eskenazi(2020)]%
        {mehri-2020-usr}
\bibfield{author}{\bibinfo{person}{Shikib Mehri} {and} \bibinfo{person}{Maxine
  Eskenazi}.} \bibinfo{year}{2020}\natexlab{}.
\newblock \showarticletitle{{USR}: An Unsupervised and Reference Free
  Evaluation Metric for Dialog Generation}. In
  \bibinfo{booktitle}{\emph{Proceedings of the 58th Annual Meeting of the
  Association for Computational Linguistics}}. \bibinfo{pages}{681--707}.
\newblock


\bibitem[Moffat et~al\mbox{.}(2015)]%
        {Moffat2015}
\bibfield{author}{\bibinfo{person}{Alistair Moffat}, \bibinfo{person}{Falk
  Scholer}, \bibinfo{person}{Paul Thomas}, {and} \bibinfo{person}{Peter
  Bailey}.} \bibinfo{year}{2015}\natexlab{}.
\newblock \showarticletitle{Pooled {Evaluation} {Over} {Query} {Variations}:
  {Users} {Are} as {Diverse} as {Systems}}. In
  \bibinfo{booktitle}{\emph{Proceedings of the 24th {ACM} {International} on
  {Conference} on {Information} and {Knowledge} {Management}}}.
  \bibinfo{pages}{1759--1762}.
\newblock


\bibitem[Moffat and Zobel(2008)]%
        {Moffat2008}
\bibfield{author}{\bibinfo{person}{Alistair Moffat} {and}
  \bibinfo{person}{Justin Zobel}.} \bibinfo{year}{2008}\natexlab{}.
\newblock \showarticletitle{Rank-Biased Precision for Measurement of Retrieval
  Effectiveness}.
\newblock \bibinfo{journal}{\emph{ACM Trans. Inf. Syst.}}  \bibinfo{volume}{27}
  (\bibinfo{year}{2008}).
\newblock


\bibitem[Oosterhuis et~al\mbox{.}(2024)]%
        {Oosterhuis:2024:RCI}
\bibfield{author}{\bibinfo{person}{Harrie Oosterhuis}, \bibinfo{person}{Rolf
  Jagerman}, \bibinfo{person}{Zhen Qin}, \bibinfo{person}{Xuanhui Wang}, {and}
  \bibinfo{person}{Michael Bendersky}.} \bibinfo{year}{2024}\natexlab{}.
\newblock \showarticletitle{Reliable Confidence Intervals for Information
  Retrieval Evaluation Using Generative A.I.}. In
  \bibinfo{booktitle}{\emph{Proceedings of the 30th ACM SIGKDD Conference on
  Knowledge Discovery and Data Mining}} \emph{(\bibinfo{series}{KDD '24})}.
  \bibinfo{pages}{2307–2317}.
\newblock


\bibitem[Padmakumar and He(2023)]%
        {padmakumar2023writing}
\bibfield{author}{\bibinfo{person}{Vishakh Padmakumar} {and}
  \bibinfo{person}{He He}.} \bibinfo{year}{2023}\natexlab{}.
\newblock \showarticletitle{Does Writing with Language Models Reduce Content
  Diversity?}
\newblock \bibinfo{journal}{\emph{arXiv preprint arXiv:2309.05196}}
  (\bibinfo{year}{2023}).
\newblock


\bibitem[Panickssery et~al\mbox{.}(2024)]%
        {panickssery2024llm}
\bibfield{author}{\bibinfo{person}{Arjun Panickssery}, \bibinfo{person}{Samuel
  Bowman}, {and} \bibinfo{person}{Shi Feng}.} \bibinfo{year}{2024}\natexlab{}.
\newblock \showarticletitle{Llm evaluators recognize and favor their own
  generations}.
\newblock \bibinfo{journal}{\emph{Advances in Neural Information Processing
  Systems}}  \bibinfo{volume}{37} (\bibinfo{year}{2024}),
  \bibinfo{pages}{68772--68802}.
\newblock


\bibitem[Parry et~al\mbox{.}(2024)]%
        {parry2024analyzing}
\bibfield{author}{\bibinfo{person}{Andrew Parry}, \bibinfo{person}{Maik
  Fr{\"o}be}, \bibinfo{person}{Sean MacAvaney}, \bibinfo{person}{Martin
  Potthast}, {and} \bibinfo{person}{Matthias Hagen}.}
  \bibinfo{year}{2024}\natexlab{}.
\newblock \showarticletitle{{Analyzing Adversarial Attacks on
  Sequence-to-Sequence Relevance Models}}. In
  \bibinfo{booktitle}{\emph{Advances in Information Retrieval. 46th European
  Conference on IR Research (ECIR 2024)}} \emph{(\bibinfo{series}{Lecture Notes
  in Computer Science})}. \bibinfo{publisher}{Springer},
  \bibinfo{address}{Berlin Heidelberg New York}.
\newblock


\bibitem[Rahmani et~al\mbox{.}(2024)]%
        {rahmani2024synthetic}
\bibfield{author}{\bibinfo{person}{Hossein~A. Rahmani}, \bibinfo{person}{Nick
  Craswell}, \bibinfo{person}{Emine Yilmaz}, \bibinfo{person}{Bhaskar Mitra},
  {and} \bibinfo{person}{Daniel Campos}.} \bibinfo{year}{2024}\natexlab{}.
\newblock \showarticletitle{{Synthetic Test Collections for Retrieval
  Evaluation}}. In \bibinfo{booktitle}{\emph{Proceedings of the 47th
  International ACM SIGIR Conference on Research and Development in Information
  Retrieval}}.
\newblock


\bibitem[Roberts et~al\mbox{.}(2020)]%
        {Roberts2020}
\bibfield{author}{\bibinfo{person}{Kirk Roberts}, \bibinfo{person}{Tasmeer
  Alam}, \bibinfo{person}{Steven Bedrick}, \bibinfo{person}{Dina
  Demner-Fushman}, \bibinfo{person}{Kyle Lo}, \bibinfo{person}{Ian Soboroff},
  \bibinfo{person}{Ellen Voorhees}, \bibinfo{person}{Lucy~Lu Wang}, {and}
  \bibinfo{person}{William~R Hersh}.} \bibinfo{year}{2020}\natexlab{}.
\newblock \showarticletitle{{TREC}-{COVID}: rationale and structure of an
  information retrieval shared task for {COVID}-19}.
\newblock \bibinfo{journal}{\emph{Journal of the American Medical Informatics
  Association}}  \bibinfo{volume}{27} (\bibinfo{year}{2020}),
  \bibinfo{pages}{1431--1436}.
\newblock


\bibitem[Roberts et~al\mbox{.}(2023)]%
        {roberts2023data}
\bibfield{author}{\bibinfo{person}{Manley Roberts}, \bibinfo{person}{Himanshu
  Thakur}, \bibinfo{person}{Christine Herlihy}, \bibinfo{person}{Colin White},
  {and} \bibinfo{person}{Samuel Dooley}.} \bibinfo{year}{2023}\natexlab{}.
\newblock \showarticletitle{Data Contamination Through the Lens of Time}.
\newblock \bibinfo{journal}{\emph{arXiv preprint arXiv:2310.10628}}
  (\bibinfo{year}{2023}).
\newblock


\bibitem[Roitero et~al\mbox{.}(2022)]%
        {Roitero2022}
\bibfield{author}{\bibinfo{person}{Kevin Roitero}, \bibinfo{person}{Alessandro
  Checco}, \bibinfo{person}{Stefano Mizzaro}, {and} \bibinfo{person}{Gianluca
  Demartini}.} \bibinfo{year}{2022}\natexlab{}.
\newblock \showarticletitle{Preferences on a {Budget}: {Prioritizing}
  {Document} {Pairs} {When} {Crowdsourcing} {Relevance} {Judgments}}. In
  \bibinfo{booktitle}{\emph{Proceedings of the {ACM} {Web} {Conference} 2022}}.
  \bibinfo{pages}{319--327}.
\newblock


\bibitem[Sainz et~al\mbox{.}(2023)]%
        {sainz2023nlp}
\bibfield{author}{\bibinfo{person}{Oscar Sainz}, \bibinfo{person}{Jon~Ander
  Campos}, \bibinfo{person}{Iker Garc{\'\i}a-Ferrero}, \bibinfo{person}{Julen
  Etxaniz}, \bibinfo{person}{Oier Lopez~de Lacalle}, {and}
  \bibinfo{person}{Eneko Agirre}.} \bibinfo{year}{2023}\natexlab{}.
\newblock \showarticletitle{NLP Evaluation in Trouble: On the Need to Measure
  LLM Data Contamination for Each Benchmark}.
\newblock \bibinfo{journal}{\emph{arXiv preprint arXiv:2310.18018}}
  (\bibinfo{year}{2023}).
\newblock


\bibitem[Sakai et~al\mbox{.}(2021)]%
        {Sakai2021}
\bibfield{author}{\bibinfo{person}{Tetsuya Sakai}, \bibinfo{person}{Sijie Tao},
  {and} \bibinfo{person}{Zhaohao Zeng}.} \bibinfo{year}{2021}\natexlab{}.
\newblock \showarticletitle{{WWW3E8}: 259,000 {Relevance} {Labels} for
  {Studying} the {Effect} of {Document} {Presentation} {Order} for {Relevance}
  {Assessors}}. In \bibinfo{booktitle}{\emph{Proceedings of the 44th
  {International} {ACM} {SIGIR} {Conference} on {Research} and {Development} in
  {Information} {Retrieval}}}. \bibinfo{pages}{2376--2382}.
\newblock


\bibitem[Scheltema(2024)]%
        {scheltema2024recommender}
\bibfield{author}{\bibinfo{person}{Nic Scheltema}.}
  \bibinfo{year}{2024}\natexlab{}.
\newblock \showarticletitle{Recommender Model Evaluation: Offline vs. Online}.
\newblock \bibinfo{journal}{\emph{Shaped Blog}} (\bibinfo{year}{2024}).
\newblock


\bibitem[Shi et~al\mbox{.}(2024)]%
        {shi2024optimization}
\bibfield{author}{\bibinfo{person}{Jiawen Shi}, \bibinfo{person}{Zenghui Yuan},
  \bibinfo{person}{Yinuo Liu}, \bibinfo{person}{Yue Huang},
  \bibinfo{person}{Pan Zhou}, \bibinfo{person}{Lichao Sun}, {and}
  \bibinfo{person}{Neil~Zhenqiang Gong}.} \bibinfo{year}{2024}\natexlab{}.
\newblock \showarticletitle{Optimization-based prompt injection attack to
  llm-as-a-judge}. In \bibinfo{booktitle}{\emph{Proceedings of the 2024 on ACM
  SIGSAC Conference on Computer and Communications Security}}.
  \bibinfo{pages}{660--674}.
\newblock


\bibitem[Shumailov et~al\mbox{.}(2023)]%
        {shumailov2023curse}
\bibfield{author}{\bibinfo{person}{Ilia Shumailov}, \bibinfo{person}{Zakhar
  Shumaylov}, \bibinfo{person}{Yiren Zhao}, \bibinfo{person}{Yarin Gal},
  \bibinfo{person}{Nicolas Papernot}, {and} \bibinfo{person}{Ross Anderson}.}
  \bibinfo{year}{2023}\natexlab{}.
\newblock \showarticletitle{The Curse of Recursion: Training on Generated Data
  Makes Models Forget}.
\newblock \bibinfo{journal}{\emph{arXiv preprint arXiv:2305.17493}}
  (\bibinfo{year}{2023}).
\newblock
\urldef\tempurl%
\url{https://arxiv.org/abs/2305.17493}
\showURL{%
\tempurl}


\bibitem[Shumailov et~al\mbox{.}(2024)]%
        {shumailov2024ai}
\bibfield{author}{\bibinfo{person}{Ilia Shumailov}, \bibinfo{person}{Zakhar
  Shumaylov}, \bibinfo{person}{Yiren Zhao}, \bibinfo{person}{Nicolas Papernot},
  \bibinfo{person}{Ross Anderson}, {and} \bibinfo{person}{Yarin Gal}.}
  \bibinfo{year}{2024}\natexlab{}.
\newblock \showarticletitle{AI models collapse when trained on recursively
  generated data}.
\newblock \bibinfo{journal}{\emph{Nature}} \bibinfo{volume}{631},
  \bibinfo{number}{8022} (\bibinfo{year}{2024}), \bibinfo{pages}{755--759}.
\newblock


\bibitem[Si et~al\mbox{.}(2024)]%
        {si2024evaluating}
\bibfield{author}{\bibinfo{person}{Chenglei Si} {et~al\mbox{.}}}
  \bibinfo{year}{2024}\natexlab{}.
\newblock \showarticletitle{Evaluating Large Language Model Biases in
  Persona-Steered Generation}. In \bibinfo{booktitle}{\emph{Findings of the
  Association for Computational Linguistics: ACL 2024}}.
  \bibinfo{pages}{6789--6800}.
\newblock


\bibitem[Singh et~al\mbox{.}(2024)]%
        {singh2024evaluation}
\bibfield{author}{\bibinfo{person}{Aaditya~K Singh},
  \bibinfo{person}{Muhammed~Yusuf Kocyigit}, \bibinfo{person}{Andrew Poulton},
  \bibinfo{person}{David Esiobu}, \bibinfo{person}{Maria Lomeli}, {and}
  \bibinfo{person}{Gergely Szilvasy}.} \bibinfo{year}{2024}\natexlab{}.
\newblock \showarticletitle{Evaluation Data Contamination in LLMs: How Do We
  Measure It and (When) Does It Matter?}
\newblock \bibinfo{journal}{\emph{arXiv preprint arXiv:2411.03923}}
  (\bibinfo{year}{2024}).
\newblock


\bibitem[Siska et~al\mbox{.}(2024)]%
        {ailem2024examining}
\bibfield{author}{\bibinfo{person}{Charlotte Siska}, \bibinfo{person}{Katerina
  Marazopoulou}, \bibinfo{person}{Melissa Ailem}, {and} \bibinfo{person}{James
  Bono}.} \bibinfo{year}{2024}\natexlab{}.
\newblock \showarticletitle{Examining the robustness of LLM evaluation to the
  distributional assumptions of benchmarks}. In
  \bibinfo{booktitle}{\emph{Proceedings of the 62nd Annual Meeting of the
  Association for Computational Linguistics (Volume 1: Long Papers)}}.
  \bibinfo{pages}{10406--10421}.
\newblock


\bibitem[Soboroff(2025)]%
        {soboroff2025don}
\bibfield{author}{\bibinfo{person}{Ian Soboroff}.}
  \bibinfo{year}{2025}\natexlab{}.
\newblock \showarticletitle{Don’t use LLMs to make relevance judgments}.
\newblock \bibinfo{journal}{\emph{Information retrieval research journal}}
  \bibinfo{volume}{1}, \bibinfo{number}{1} (\bibinfo{year}{2025}),
  \bibinfo{pages}{10--54195}.
\newblock


\bibitem[Spärck~Jones and van Rijsbergen(1975)]%
        {spark1975report}
\bibfield{author}{\bibinfo{person}{Karen Spärck~Jones} {and}
  \bibinfo{person}{C.~J. van Rijsbergen}.} \bibinfo{year}{1975}\natexlab{}.
\newblock \showarticletitle{Report on the need for and provision of an `ideal'
  information retrieval test collection}.
\newblock \bibinfo{journal}{\emph{Computer Laboratory}} (\bibinfo{year}{1975}).
\newblock


\bibitem[Steyvers et~al\mbox{.}(2025)]%
        {steyvers2025large}
\bibfield{author}{\bibinfo{person}{Mark Steyvers}, \bibinfo{person}{Heliodoro
  Tejeda}, \bibinfo{person}{Aakriti Kumar}, \bibinfo{person}{Catarina Belem},
  \bibinfo{person}{Sheer Karny}, \bibinfo{person}{Xinyue Hu},
  \bibinfo{person}{Lukas~W Mayer}, {and} \bibinfo{person}{Padhraic Smyth}.}
  \bibinfo{year}{2025}\natexlab{}.
\newblock \showarticletitle{What large language models know and what people
  think they know}.
\newblock \bibinfo{journal}{\emph{Nature Machine Intelligence}}
  (\bibinfo{year}{2025}), \bibinfo{pages}{1--11}.
\newblock


\bibitem[Thomas et~al\mbox{.}(2024a)]%
        {thomas2024matters}
\bibfield{author}{\bibinfo{person}{Paul Thomas}, \bibinfo{person}{Gabriella
  Kazai}, \bibinfo{person}{Nick Craswell}, {and} \bibinfo{person}{Seth
  Spielman}.} \bibinfo{year}{2024}\natexlab{a}.
\newblock \showarticletitle{What matters in a measure? A perspective from
  large-scale search evaluation}. In \bibinfo{booktitle}{\emph{Proceedings of
  the 47th International ACM SIGIR Conference on Research and Development in
  Information Retrieval}}. \bibinfo{pages}{282--292}.
\newblock


\bibitem[Thomas et~al\mbox{.}(2024b)]%
        {thomas2024large}
\bibfield{author}{\bibinfo{person}{Paul Thomas}, \bibinfo{person}{Seth
  Spielman}, \bibinfo{person}{Nick Craswell}, {and} \bibinfo{person}{Bhaskar
  Mitra}.} \bibinfo{year}{2024}\natexlab{b}.
\newblock \showarticletitle{{Large Language Models can Accurately Predict
  Searcher Preferences}}. In \bibinfo{booktitle}{\emph{Proceedings of the 47th
  International ACM SIGIR Conference on Research and Development in Information
  Retrieval}}.
\newblock


\bibitem[Tseng et~al\mbox{.}(2024)]%
        {tseng2024two}
\bibfield{author}{\bibinfo{person}{Yu-Min Tseng}, \bibinfo{person}{Yu-Chao
  Huang}, \bibinfo{person}{Teng-Yun Hsiao}, \bibinfo{person}{Wei-Lin Chen},
  \bibinfo{person}{Chao-Wei Huang}, \bibinfo{person}{Yu Meng}, {and}
  \bibinfo{person}{Yun-Nung Chen}.} \bibinfo{year}{2024}\natexlab{}.
\newblock \showarticletitle{Two Tales of Persona in LLMs: A Survey of
  Role-Playing and Personalization}. In \bibinfo{booktitle}{\emph{Findings of
  the Association for Computational Linguistics: EMNLP 2024}}.
  \bibinfo{pages}{16612--16631}.
\newblock


\bibitem[Upadhyay et~al\mbox{.}(2024a)]%
        {upadhyay2024llm}
\bibfield{author}{\bibinfo{person}{Shivani Upadhyay}, \bibinfo{person}{Ronak
  Pradeep}, \bibinfo{person}{Nandan Thakur}, \bibinfo{person}{Daniel Campos},
  \bibinfo{person}{Nick Craswell}, \bibinfo{person}{Ian Soboroff},
  \bibinfo{person}{Hoa~Trang Dang}, {and} \bibinfo{person}{Jimmy Lin}.}
  \bibinfo{year}{2024}\natexlab{a}.
\newblock \showarticletitle{A Large-Scale Study of Relevance Assessments with
  Large Language Models: An Initial Look}.
\newblock \bibinfo{journal}{\emph{arXiv preprint arXiv:2411.08275}}
  (\bibinfo{year}{2024}).
\newblock


\bibitem[Upadhyay et~al\mbox{.}(2024b)]%
        {upadhyay2024large}
\bibfield{author}{\bibinfo{person}{Shivani Upadhyay}, \bibinfo{person}{Ronak
  Pradeep}, \bibinfo{person}{Nandan Thakur}, \bibinfo{person}{Daniel Campos},
  \bibinfo{person}{Nick Craswell}, \bibinfo{person}{Ian Soboroff},
  \bibinfo{person}{Hoa~Trang Dang}, {and} \bibinfo{person}{Jimmy Lin}.}
  \bibinfo{year}{2024}\natexlab{b}.
\newblock \showarticletitle{A Large-Scale Study of Relevance Assessments with
  Large Language Models: An Initial Look}.
\newblock \bibinfo{journal}{\emph{arXiv preprint arXiv:2411.08275}}
  (\bibinfo{year}{2024}).
\newblock


\bibitem[Upadhyay et~al\mbox{.}(2024c)]%
        {upadhyay2024umbrela}
\bibfield{author}{\bibinfo{person}{Shivani Upadhyay}, \bibinfo{person}{Ronak
  Pradeep}, \bibinfo{person}{Nandan Thakur}, \bibinfo{person}{Nick Craswell},
  {and} \bibinfo{person}{Jimmy Lin}.} \bibinfo{year}{2024}\natexlab{c}.
\newblock \showarticletitle{{UMBRELA: UMbrela is the (Open-Source Reproduction
  of the) Bing RELevance Assessor}}.
\newblock \bibinfo{journal}{\emph{arXiv preprint arXiv:2406.06519}}
  (\bibinfo{year}{2024}).
\newblock


\bibitem[Voorhees(2019)]%
        {Voorhees2019}
\bibfield{author}{\bibinfo{person}{Ellen~M. Voorhees}.}
  \bibinfo{year}{2019}\natexlab{}.
\newblock \showarticletitle{The Evolution of Cranfield}.
\newblock In \bibinfo{booktitle}{\emph{Information Retrieval Evaluation in a
  Changing World}}, \bibfield{editor}{\bibinfo{person}{Nicola Ferro} {and}
  \bibinfo{person}{Carol Peters}} (Eds.). Vol.~\bibinfo{volume}{41}.
  \bibinfo{publisher}{Springer International Publishing},
  \bibinfo{pages}{45--69}.
\newblock


\bibitem[Voorhees et~al\mbox{.}(2022)]%
        {voorhees2022can}
\bibfield{author}{\bibinfo{person}{Ellen~M Voorhees}, \bibinfo{person}{Ian
  Soboroff}, {and} \bibinfo{person}{Jimmy Lin}.}
  \bibinfo{year}{2022}\natexlab{}.
\newblock \showarticletitle{Can Old TREC Collections Reliably Evaluate Modern
  Neural Retrieval Models?}
\newblock \bibinfo{journal}{\emph{arXiv preprint arXiv:2201.11086}}
  (\bibinfo{year}{2022}).
\newblock


\bibitem[Wang et~al\mbox{.}(2024)]%
        {wang2024}
\bibfield{author}{\bibinfo{person}{Liang Wang}, \bibinfo{person}{Nan Yang},
  \bibinfo{person}{Xiaolong Huang}, \bibinfo{person}{Binxing Jiao},
  \bibinfo{person}{Linjun Yang}, \bibinfo{person}{Daxin Jiang},
  \bibinfo{person}{Rangan Majumder}, {and} \bibinfo{person}{Furu Wei}.}
  \bibinfo{year}{2024}\natexlab{}.
\newblock \showarticletitle{Text Embeddings by Weakly-Supervised Contrastive
  Pre-training}.
\newblock \bibinfo{journal}{\emph{arXiv preprint arXiv:2212.03533}}
  (\bibinfo{year}{2024}).
\newblock


\bibitem[Wei et~al\mbox{.}(2024)]%
        {wei2024rethinking}
\bibfield{author}{\bibinfo{person}{Fangyun Wei}, \bibinfo{person}{Xi Chen},
  {and} \bibinfo{person}{Lin Luo}.} \bibinfo{year}{2024}\natexlab{}.
\newblock \showarticletitle{Rethinking generative large language model
  evaluation for semantic comprehension}.
\newblock \bibinfo{journal}{\emph{arXiv preprint arXiv:2403.07872}}
  (\bibinfo{year}{2024}).
\newblock


\bibitem[Xu et~al\mbox{.}(2024a)]%
        {xu2024benchmark}
\bibfield{author}{\bibinfo{person}{Cheng Xu}, \bibinfo{person}{Shuhao Guan},
  \bibinfo{person}{Derek Greene}, \bibinfo{person}{M Kechadi}, {et~al\mbox{.}}}
  \bibinfo{year}{2024}\natexlab{a}.
\newblock \showarticletitle{Benchmark data contamination of large language
  models: A survey}.
\newblock \bibinfo{journal}{\emph{arXiv preprint arXiv:2406.04244}}
  (\bibinfo{year}{2024}).
\newblock


\bibitem[Xu et~al\mbox{.}(2024b)]%
        {xu2024pride}
\bibfield{author}{\bibinfo{person}{Wenda Xu}, \bibinfo{person}{Guanglei Zhu},
  \bibinfo{person}{Xuandong Zhao}, \bibinfo{person}{Liangming Pan},
  \bibinfo{person}{Lei Li}, {and} \bibinfo{person}{William Wang}.}
  \bibinfo{year}{2024}\natexlab{b}.
\newblock \showarticletitle{Pride and Prejudice: LLM Amplifies Self-Bias in
  Self-Refinement}. In \bibinfo{booktitle}{\emph{Proceedings of the 62nd Annual
  Meeting of the Association for Computational Linguistics (Volume 1: Long
  Papers)}}. \bibinfo{pages}{15474--15492}.
\newblock


\bibitem[Yeo et~al\mbox{.}(2024)]%
        {yeo2024selftraining}
\bibfield{author}{\bibinfo{person}{Wei~Jie Yeo}, \bibinfo{person}{Teddy
  Ferdinan}, \bibinfo{person}{Przemyslaw Kazienko}, \bibinfo{person}{Ranjan
  Satapathy}, {and} \bibinfo{person}{Erik Cambria}.}
  \bibinfo{year}{2024}\natexlab{}.
\newblock \showarticletitle{Self-training Large Language Models through
  Knowledge Detection}. In \bibinfo{booktitle}{\emph{Findings of the
  Association for Computational Linguistics: EMNLP 2024}}.
  \bibinfo{pages}{12345--12356}.
\newblock


\bibitem[Zhang et~al\mbox{.}(2020a)]%
        {zhang2020better}
\bibfield{author}{\bibinfo{person}{Fan Zhang} {et~al\mbox{.}}}
  \bibinfo{year}{2020}\natexlab{a}.
\newblock \showarticletitle{Towards a Better Understanding of Evaluation
  Metrics}. In \bibinfo{booktitle}{\emph{Proceedings of the 43rd International
  ACM SIGIR Conference on Research and Development in Information Retrieval}}.
  \bibinfo{pages}{1439--1442}.
\newblock


\bibitem[Zhang et~al\mbox{.}(2023)]%
        {zhang2023constructing}
\bibfield{author}{\bibinfo{person}{Fan Zhang} {et~al\mbox{.}}}
  \bibinfo{year}{2023}\natexlab{}.
\newblock \showarticletitle{Constructing and Meta-Evaluating State-Aware
  Evaluation Metrics for Information Retrieval}.
\newblock \bibinfo{journal}{\emph{Information Retrieval Journal}}
  (\bibinfo{year}{2023}).
\newblock


\bibitem[Zhang et~al\mbox{.}(2020b)]%
        {zhang2020models}
\bibfield{author}{\bibinfo{person}{Fan Zhang}, \bibinfo{person}{Jiaxin Mao},
  \bibinfo{person}{Yiqun Liu}, \bibinfo{person}{Xiaohui Xie},
  \bibinfo{person}{Weizhi Ma}, \bibinfo{person}{Min Zhang}, {and}
  \bibinfo{person}{Shaoping Ma}.} \bibinfo{year}{2020}\natexlab{b}.
\newblock \showarticletitle{Models versus satisfaction: Towards a better
  understanding of evaluation metrics}. In
  \bibinfo{booktitle}{\emph{Proceedings of the 43rd international acm sigir
  conference on research and development in information retrieval}}.
  \bibinfo{pages}{379--388}.
\newblock


\bibitem[Zhou et~al\mbox{.}(2023)]%
        {zhou2023dont}
\bibfield{author}{\bibinfo{person}{Kun Zhou}, \bibinfo{person}{Yutao Zhu},
  \bibinfo{person}{Zhipeng Chen}, \bibinfo{person}{Wentong Chen},
  \bibinfo{person}{Wayne~Xin Zhao}, \bibinfo{person}{Xu Chen},
  \bibinfo{person}{Yankai Lin}, \bibinfo{person}{Ji-Rong Wen}, {and}
  \bibinfo{person}{Jiawei Han}.} \bibinfo{year}{2023}\natexlab{}.
\newblock \showarticletitle{Don't Make Your LLM an Evaluation Benchmark
  Cheater}.
\newblock \bibinfo{journal}{\emph{arXiv preprint arXiv:2311.01964}}
  (\bibinfo{year}{2023}).
\newblock


\bibitem[Zhou et~al\mbox{.}(2025)]%
        {zhou2025evaluating}
\bibfield{author}{\bibinfo{person}{Yilun Zhou}, \bibinfo{person}{Austin Xu},
  \bibinfo{person}{Peifeng Wang}, \bibinfo{person}{Caiming Xiong}, {and}
  \bibinfo{person}{Shafiq Joty}.} \bibinfo{year}{2025}\natexlab{}.
\newblock \showarticletitle{Evaluating Judges as Evaluators: The JETTS
  Benchmark of LLM-as-Judges as Test-Time Scaling Evaluators}.
\newblock \bibinfo{journal}{\emph{arXiv preprint arXiv:2504.15253}}
  (\bibinfo{year}{2025}).
\newblock


\end{thebibliography}
